\documentclass[pra,twocolumn,nofootinbib]{revtex4-1}
\usepackage{dcolumn}
\usepackage{bm}
\usepackage{amsmath,amssymb}
\usepackage{bm}
\usepackage[dvipdfmx]{graphicx} 
\usepackage{ascmac}
\usepackage[latin1]{inputenc}
\usepackage{braket}
\usepackage{txfonts}
\usepackage{mathrsfs}
\usepackage{color}
\usepackage[bottom]{footmisc}
\usepackage{here}
\usepackage{ulem}
\def\be{\begin{equation}}  
\def\ee{\end{equation}}     
\def\I{\mathrm{i}}  
\def\tr{\mathrm{tr}}
\def\e{\mathrm{e}} 

\def\p{| \vec{p} \, |}
\def\v{V}
\def\sfV{\mathsf{V}}

\def\kp{\kappa^\prime} 
\def\JS{J_{\mathrm{S}}}
\def\JR{J_{\mathrm{R}}}

\def\erfc{\mathrm{erfc}} 
\begin{document}
\title{Error tradeoff relation for estimating the unitary-shift parameter of a relativistic spin-1/2 particle}
\author{Shin Funada}
\author{Jun Suzuki}
\affiliation{%
Graduate School of Informatics and Engineering, The University of Electro-Communications, 
1-5-1 Chofugaoka, Chofu-shi, Tokyo, 182-8585 Japan
}%

\date{\today}

\begin{abstract}
The purpose of this paper is to discuss the existence of a nontrivial tradeoff relation 
for estimating two unitary-shift parameters in a relativistic spin-1/2 system. 
It is shown that any moving observer cannot estimate two parameters simultaneously, 
even though a parametric model is classical in the rest frame. 
This transition from the classical model to a genuine quantum model is 
investigated analytically using a one-parameter family of quantum Fisher information matrices. 
This paper proposes to use an indicator that can not only detect the existence of a tradeoff relation 
but can also evaluate its strength. 
Based on the proposed indicator, this paper investigates the nature of the tradeoff relation in detail. 

\end{abstract}

\maketitle 
\section{Introduction}
%

Incompatibility for estimating multiple parameters is one of the intrinsic natures of quantum estimation theory \cite{helstrom,holevo,hayashi_ws}. 
Any two parameters encoded in a quantum state cannot be estimated simultaneously with the minimum estimation error, and hence one faces to the tradeoff relation. 
This is because an optimal measurement for one parameter prohibits us to estimate 
the other parameter with the best estimation accuracy and vice verse. 
In the worst case, the optimal measurement for one parameter could not estimate the other at all. 
This extreme case is connected to the uncertainty relation formulated in the framework of quantum estimation theory 
\cite{holevo,gibilisco,watanabe,gzjfw2016,kull}. 
Up to now, incompatibility has been addressed in various models ranging from abstract mathematical models to physical models 
\cite{carollo,zhu,lu,belliardo,razavian,candeloro,huang,asjad,chen}. 

Relativistic quantum estimation is a relatively new topic in quantum estimation theory, 
although the pioneering work \cite{braunstein96} appeared about a quarter century ago. 
In these studies, relativistic metrology has been shown to be a potential resource for quantum advantage due to relativistic quantum theory \cite{ahmadi2,ahmadi,tian,liu}. 
It seems, however, incompatibility in relativistic estimation has not been explored so far. 
In particular, there is no previous study that investigates how incompatibility changes for different moving observers. 
This paper's main achievement is to progress in this line of research. 

We take a specific model for a relativistic spin-1/2 particle which was studied in Ref.~\cite{sf}. 
In this model, we set up a classical-like Gaussian wave packet in the rest frame, 
and we wish to estimate the unitary-shift parameters for the $x,y$ directions. 
Since there is no correlation between the two directions, one can perform a precise position measurement 
for each direction.  
Hence, there exists no tradeoff relation that gives rise to incompatibility upon estimating two different parameters. 
However, a moving observer along the $z$-direction sees this state distorted due to the Wigner rotation \cite{weinberg,halpern}. 
When the observer accesses the position degrees of freedom only, the reduced state becomes a mixed state 
due to the information loss regarding the spin of the particle. 
In Ref.~\cite{sf}, we showed this model does not satisfy the so-called weak-commutativity condition \cite{ragy,suzuki3}. 
Therefore, one cannot estimate two parameters simultaneously even in the asymptotic limit. 
Yet, we could not discuss a tradeoff relation since we only analyzed the symmetric logarithmic derivatives (SLD) Cram\'er-Rao (CR) bound. 

In this paper, we propose to use one-parameter family of quantum CR bounds to evaluate the tradeoff relation 
between two diagonal components for the mean-square-error (MSE) matrix \cite{suzuki_ld,yamagata}. 
By combining the SLD CR bound and one-parameter family of quantum CR bounds, 
we define an indicator of the existence of the tradeoff relation. 
If this indicator is positive, we definitely conclude that there is a tradeoff relation.  
In this way, the proposed indicator can witness the existence of a tradeoff relation. 
Furthermore, the value of the indicator corresponds to the strength of the tradeoff relation and hence it serves more than just a witness. 
We apply this incompatibility witness to our relativistic model, 
and we show that any moving observer cannot estimate two parameters simultaneously. 
In other words, incompatibility is inevitable no matter how slow the observer is. 

The outline of this paper is as follows. 
In Sec.~II, we give a physical mode in the rest frame. Next, we derive the momentum representation of 
the wave function in the moving frame. We also give parametric models in the rest frame and the moving frame. 
Section III introduces an indicator of tradeoff relation $\omega$.  
We show that we can conclude that there exists a tradeoff relation when the indicator $\omega$ is positive. 
Next, we show that the indicator $\omega$ is always positive when the observer's velocity is nonzero 
if the $\lambda$ in the $\lambda$LD Fisher information matrix is in an appropriate range. 
Section IV and Sec.~V give a discussion and conclusion, respectively. 
Appendices give supplemental information for the calculations in detail. 
\section{Model and $\lambda$LD Fisher information matrix}
In our previous study~\cite{sf}, we investigated a two-parameter unitary-shift model of a relativistic spin-1/2 particle. We analyzed the estimation accuracy limit by a lower bound based on the SLD CR bound. 
Because the SLD CR bound is not attainable due to incompatibility, we could not address the question of a tradeoff relation in full detail. 
To proceed with further discussion, we employ the idea of combining two different quantum CR bounds, the SLD and right logarithmic derivative (RLD) CR bounds, which was proposed in Ref.~\cite{sf2}. 
The previous method does not work for this model since the RLD Fisher information matrix does not exist for our model. 
In this paper, we extend this method by using another type of quantum Fisher information matrix called  
$\lambda$LD Fisher information matrix~\cite{suzuki_ld,yamagata}.
We will show that we can detect and discuss the strength of a tradeoff between estimation error for two different parameters. 

\subsection{Physical model} \label{sec:Model}
This section briefly summarizes a physical model used in our previous study. 
See~\cite{sf} for more in detail. 
Consider a spin-1/2 relativistic particle with the rest mass $m$ and assume that its spin is down in the rest frame. 
The wave function is set as an isotropic Gaussian function with the spread $\kappa$ in the $x,y$ direction, 
and a plane wave function is chosen for the $z$-direction. 
To discuss a relativistic effect, consider an observer who is moving along the $z$ axis with a constant velocity $\v$. 
To give the largest relativistic effect, the $z$ direction is chosen for the observer's motion ~\cite{terashima}.  
Natural units, i.e., $\hbar=1$ and $c=1$ will be used unless otherwise stated. 

To apply the Wigner rotation later, it is convenient to describe the particle in the momentum representation. 
The state vector in the rest frame is given by
\begin{align}
\ket{\Psi_\downarrow}&= \int d^3p \, \varphi_0(p^1) \varphi_0(p^2)\delta(p^3) \ket{\vec{p}, \downarrow} \label{eq:Psi0},
\end{align}
where $\delta(p^{3})$ denotes the Dirac delta function and $\varphi_0(t)$ is the Gaussian function 
\be
\varphi_0(t)=
\frac{\kappa^{1/2}}{\pi^{1/4}}\, \e^{- \frac12\kappa^2 t^2}. \label{eq:phi_0}
\ee
We remind that the $\kappa$ is the spread in the coordinate representation. 
In the above expression \eqref{eq:Psi0}, we denote the spatial part of the four-momentum vector by $\vec{p}$, 
that is $\vec{p}=(p^1,p^2,p^3)$. 
%
\subsection{Parametric model: rest frame}
A two-parameter unitary-shift model in the rest frame is defined as follows. 
Define a unitary transformation generated by the momentum operators in the $x$ and $y$ direction, $\hat{p}^1$ and $\hat{p}^2$, by 
\be
U(\theta)=\e^{-\I \hat{p}^1 \theta_1 - \I \hat{p}^2 \theta_2}. \label{eq:unitary}
\ee
This unitary-shift operator encodes two parameters $(\theta_1,\theta_2)$ in the $x,y$ direction of the wave function. 
By applying $U(\theta)$ to the state vector $\ket{\Psi_\downarrow}$, we define the pure state model: 
\be\label{eq:model_rest}
\mathcal{M}_{\rm rest}= \big\{ \rho_\theta \,\big|\, \theta = (\theta_1, \theta_2) \subset \mathbb{R}^2 \big\}, 
\ee
where $\rho_\theta$ is 
\be
\rho_\theta
=U(\theta) \ket{\Psi_\downarrow} \bra{\Psi_\downarrow}U^\dagger (\theta). \label{eq:rho_theta2}
\ee
%
\subsection{State in the moving frame}
We next discuss the state vector in the moving frame. 
In our model, the state vector in the rest frame is in a spin-down state, $\ket{\Psi_{\downarrow}(\theta)}=U(\theta) \ket{\Psi_\downarrow}$. 
The state vector in the moving frame is given by the Wigner rotation~\cite{halpern,weinberg} $U(\Lambda)$ with $\Lambda$ the Lorentz transformation on the state as 
\be \label{eq:Psilambda}
\ket{\Psi^\Lambda (\theta)} =U(\Lambda) \ket{\Psi_\downarrow (\theta)}
= \sum_{\sigma=\downarrow, \uparrow} \ket{\psi^\Lambda_{\: \sigma}(\theta)} \ket{\sigma},  
\ee
where $\ket{\psi^\Lambda_{\: \sigma}(\theta)}$ are 
\begin{align}
 \ket{\psi^\Lambda_{\: \sigma}(\theta)}&= \int d^3p \sqrt{\frac{(\Lambda p)^0}{p^0}} F_{\theta, \, \sigma}({p}^1, {p}^2) \delta(p^3) \ket{\Lambda \vec{p}}, 
  \label{eq:Psi_pi}\\
F_{\theta, \, \downarrow}(p^1, p^2)
&=\varphi_0(p^1)\varphi_0(p^2) \e^{- \I p^1 \theta_1- \I p^2 \theta_2} \cos \frac{\alpha(\p)}{2}, \label{eq:F1} \\
F_{\theta, \, \uparrow}(p^1, p^2)
&=-\varphi_0(p^1)\varphi_0(p^2)  \e^{- \I p^1 \theta_1- \I p^2 \theta_2} \e^{\I \phi(p^1, \, p^2)}\sin \frac{\alpha(\p)}{2}, \label{eq:F2} \\
\p &=\sqrt{(p^1)^2+(p^2)^2}, \nonumber \\
\e^{\I \phi(p^1, \, p^2)}&= \frac{p^1}{\p}+ \I \frac{p^2}{\p}, \nonumber \\
\cos \alpha(\p)&=\frac{\sqrt{m^2+ \p^2} + m \cosh \chi}{\sqrt{m^2+ \p^2} \cosh \chi + m }, \label{eq:cosbeta} \\
\sin \alpha(\p)&= - \frac{\p \sinh{\chi}}{\sqrt{m^2+ \p^2} \cosh \chi + m },  \label{eq:sinbeta}\\
\chi&=\tanh^{-1}(V). 
\end{align}

When the observer's velocity $V$ is not zero, i.e., when the observer is moving, 
$F_{\theta, \, \uparrow}(p^1, p^2) \neq 0$ holds. Therefore, 
the particle spin `rotate'.  
We give a brief summary of the derivations of Eqs.~\eqref{eq:Psilambda},~\eqref{eq:Psi_pi},~\eqref{eq:F1}, and~\eqref{eq:F2} 
in~Appendix~\ref{sec:Wigner} and the full detail account is given in Ref.~\cite{sf}. 

\subsection{Parametric model: Moving frame}
The parametric model defined by the state vector in the moving frame 
$\ket{\Psi^\Lambda (\theta)} =U(\Lambda) \ket{\Psi_\downarrow (\theta)}$ is unitary equivalent to the model in the rest frame. Hence, all the properties remain the same. 
To introduce a relativistic effect of the Wigner rotation on the model, 
we take the partial trace over the spin degree of freedom \cite{sf}. 
This corresponds to the situation where the moving observer accesses the position degree of freedom only. 
The parametric model is then defined as 
\be \label{eq:model_rel}
\mathcal{M}^{\Lambda}= \big\{  \rho^\Lambda(\theta) \, \big| \, \theta=(\theta_1, \theta_2) \in \mathbb{R}^2 \big\}, 
\ee
where
\begin{align}
 \rho^\Lambda(\theta)&=\tr_\sigma \ket{\Psi^\Lambda(\theta)} \bra{\Psi^\Lambda(\theta)}\nonumber\\
 &=\ket{{\psi}^\Lambda_{\: \downarrow}(\theta)}\bra{{\psi}^\Lambda_{\: \downarrow}(\theta)}
+\ket{{\psi}^\Lambda_{\: \uparrow}(\theta)}\bra{{\psi}^\Lambda_{\: \uparrow}(\theta)}. 
\label{eq:wavefunction2}
\end{align}
It is worth reminding ourselves that the vectors $\ket{{\psi}^\Lambda_{\: \sigma}(\theta)}$ are not normalized. 
Their normalized vectors will be denoted by $\ket{\bar{\psi}^\Lambda_{\: \sigma}(\theta)}$. 
We have a remark for this model. 
First, the parametric state in the model \eqref{eq:model_rel} is not full rank.
Thus, the RLDs do not exist in this singular
model. 
Next, the state vectors $\ket{\psi^\Lambda_{\: \uparrow}(\theta)}$ and $\ket{\psi^\Lambda_{\: \downarrow}(\theta)}$ are 
orthogonal~\cite{sf}, i.e., $\braket{\psi^\Lambda_{\: \uparrow}(\theta) | \psi^\Lambda_{\: \downarrow}(\theta)}=0$. 
This model is expressed by the statistical mixture of the two orthogonal state vectors; hence, it is a rank 2 model. 

\begin{figure}[t]
\begin{center}
\includegraphics[width=8cm]{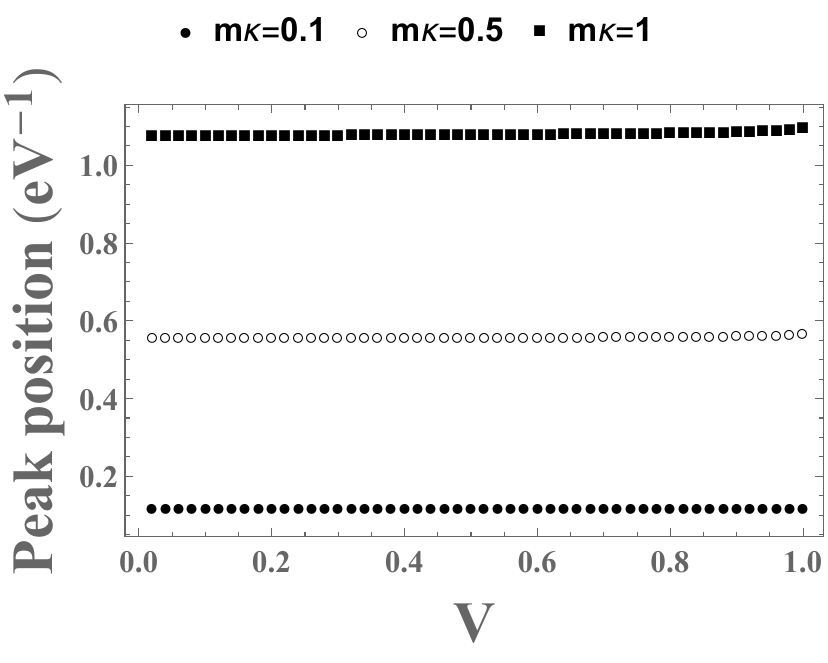}
\caption{Numerically calculated peak position at $z=0$ ($x^3=0$) as a function of the observer's velocity $V$ 
at $m \kappa$=0.1, 0.5, and 1. The $m$ and $\kappa$ are the rest mass of the particle 
and the spread of the wave function, respectively. 
The velocity $V$ and $m \kappa$ are dimensionless in the natural units. 
The peak is derived from the first derivative of the probability density. 
The length has the dimension of the inverse of the energy in the natural units. When $m \kappa =1$, for example, 
the spread $\kappa$ is equal to the Compton wavelength of the particle.}
\label{fig_peak}
\end{center}
\end{figure}

To get a physical insight into the model, let us briefly discuss the probability density of the spin-up state in the coordinate representation. 
As for the probability density of the spin-up state for the nonzero observer's velocity, i.e., $V \neq 0$, 
we can show that $ |\braket{x | \bar{\psi}^\Lambda_{\: \uparrow}(\theta)} |^2$ has rotational symmetry around 
the center of the original Gaussian wave packet. 
If $ V \neq 0$, we have two peaks in the particle's probability density, 
this addition of the other peak makes it difficult to make a position estimation~\cite{sf}. 
See Appendix~\ref{sec:peak_position} for the derivation. 
Figure~\ref{fig_peak} shows the peak position of the spin up state, $ |\braket{x | \bar{\psi}^\Lambda_{\: \uparrow}(\theta=0)} |$ from the $z$-axis as a function of the observer's velocity $V$. 
The numerical calculation indicates that the peak position is about the same as $m \kappa$. For the same particle, the peak position is 
determined by the spread $\kappa$. 

\subsection{$\lambda$LD Fisher information matrix}
Let us quickly review the $\lambda$LD and the $\lambda$LD Fisher information matrix \cite{suzuki_ld,yamagata}.
The $\lambda$LD is a one-parameter family of logarithmic derivatives, which is defined by a solution of the following equation: 
\begin{align}
\frac{\partial \rho_\theta }{\partial \theta_j}
&= \frac{1+ \lambda}{2} \rho_\theta L_{\lambda, j}( \theta)
+ \frac{1- \lambda}{2}   L_{\lambda, j}(\theta) \rho_\theta, 
\end{align}
where $-1 \leq \lambda \leq 1$. 
By using the $\lambda$LD, 
the $(i,j)$ component of the $\lambda$LD Fisher information matrix $J_{\lambda}(\theta)$ is defined by
\begin{align}
J_{\lambda,ij}(\theta)
&=\frac{1+\lambda}{2} \tr \big(\rho_\theta L_{\lambda, j}( \theta) L^\dagger_{\lambda, i}(\theta) \big)+
 \frac{1-\lambda}{2} \tr \big(\rho_\theta L^\dagger_{\lambda, i}( \theta) L_{\lambda, j}(\theta) \big). 
\label{eq:def_lambdaLD_FIM}
 \end{align}
By definition, we see that $\lambda=0$ corresponds to the SLD and $\lambda=1$ does to the RLD. 
(Cf. $\lambda=-1$ is so called left logarithmic derivative.) 
As the model $\mathcal{M}^{\Lambda}$ is not full rank, 
the $\lambda$LD is not uniquely defined. 
However, the $\lambda$LD Fisher information matrix is uniquely defined. 
A derivation of the $\lambda$LD and $\lambda$LD Fisher information 
matrix for a rank deficient model is given in Appendix~\ref{ch7_sec:lambdaLD_no_full_rank}. 

With using the formula Eq.~\eqref{app_eq:Jlambda_formula} given in Appendix~\ref{ch7_sec:lambdaLD_no_full_rank}, we can calculate the 
$\lambda$LD Fisher information matrix for the model~\eqref{eq:model_rel}. 
Since the model is still unitary, there is not parametric dependence on the quantum Fisher information matrix. 
In the following discussion, we will omit the parameter $\theta$. 

The $\mathrm{\lambda LD}$ Fisher information matrix $J_{\lambda}$ is obtained as
\be
J_{\lambda}= \frac{2 }{\kappa^2 (1- \lambda^2)(1 -\lambda^2 \xi^2)} 
\begin{pmatrix}
1-\zeta ^2 -\lambda^2 \xi^2 & 
- \I\lambda \zeta^2  \xi  \\
 \I \lambda\zeta^2  \xi  &
1-\zeta^2 -\lambda^2 \xi^2 \\
\end{pmatrix}, \label{eq:Jlambda}
\ee
where $\zeta$ and $\xi$ are defined by \footnote{In Ref.~\cite{sf}, the integral $\zeta$ was defined as a different symbol $\eta$. The correspondence to the current paper is $\zeta=\sqrt{2}\kappa\eta$.}
\begin{align}
\zeta&={\sqrt{2}}m^3 {\kappa}^3 \v \int^\infty_0 dt \frac{  t^3 }{\sqrt{1+t^2} 
 + \sqrt{1- {\v}^2}}\e^{-m^2{\kappa}^2 t^2}, \label{eq:zeta} \\
\xi &=
m^2 {\kappa}^2 \int^\infty_0 dt \frac{ 2 t (1+\sqrt{1+t^2} \sqrt{1- {\v}^2})}{\sqrt{1+t^2} 
 + \sqrt{1- {\v}^2}}\e^{-m^2 \kappa^2 t^2}. \label{eq:xi}
\end{align}

The inverse of the $\lambda$LD Fisher information matrix, $J^{-1}_{\lambda}$ is given by
\begin{align}
J_{\lambda}^{\: -1}&= 
\frac{\kappa^2}{2} \frac{1-\lambda ^2 }{(1-\zeta ^2)^2 -\lambda ^2 \xi ^2} 
\begin{pmatrix}
1-\zeta^2 -\lambda ^2 \xi ^2 & 
\I \lambda\zeta^2  \xi  \\
- \I \lambda\zeta^2  \xi  &
1-\zeta^2 -\lambda ^2 \xi ^2 \\
\end{pmatrix}. \label{ch7_eq:Jlambdainv}
\end{align}
In the following, we express the $(i, j)$ component of 
the inverse of the $\lambda$LD Fisher information matrix, $J_{\lambda}^{\: -1}$ as
\be
[J_{\lambda}^{\: -1}]_{i j}=J^{-1}_{\: \lambda , ij}.
\ee
We obtain the inverse of the SLD and the RLD Fisher information matrices, 
$\JS^{\:-1}$ and $\JR^{\:-1}$ when $\lambda$=0 and 1, respectively. 
We see that $\JR^{\:-1}$ is a zero matrix, i.e., the RLD CR inequality gives a trivial bound. 
The inverse of the SLD Fisher information matrix $\JS^{\:-1}$ is obtained as 
\begin{align}
\JS^{\: -1}&= \frac{\kappa^2}{2(1-\zeta^2)}
\begin{pmatrix}
1 & 0 \\
0 & 1 \\
\end{pmatrix}. \label{eq:SLD_Fisher} 
\end{align}
This recovers the result of Ref.~\cite{sf}. 
Lastly, we observe that the $\lambda$LD Fisher information matrix is symmetric 
with respect to two parameters $\theta_1$ and $\theta_2$. 
This is because of rotational symmetry in the parametric model. 

\section{Tradeoff relation for estimation error of unitary-shift parameters}
\subsection{Indicator of tradeoff relation}
In order to discuss a tradeoff relation, we introduce an indicator of tradeoff relation $\omega$ 
derived from our way of quantifying a tradeoff relation. 
The $\omega$ is a quantity that not only guarantees the existence of a tradeoff relationship 
but also quantifies the strength of it. 
In particular, if $\omega$ is positive, we can 
conclude the existence of a tradeoff relationship. 
Further, the larger the positive value for the $\omega$ is, the more significant the tradeoff relation becomes. 

\begin{figure}[t]
\begin{center}
\includegraphics[width=9cm]{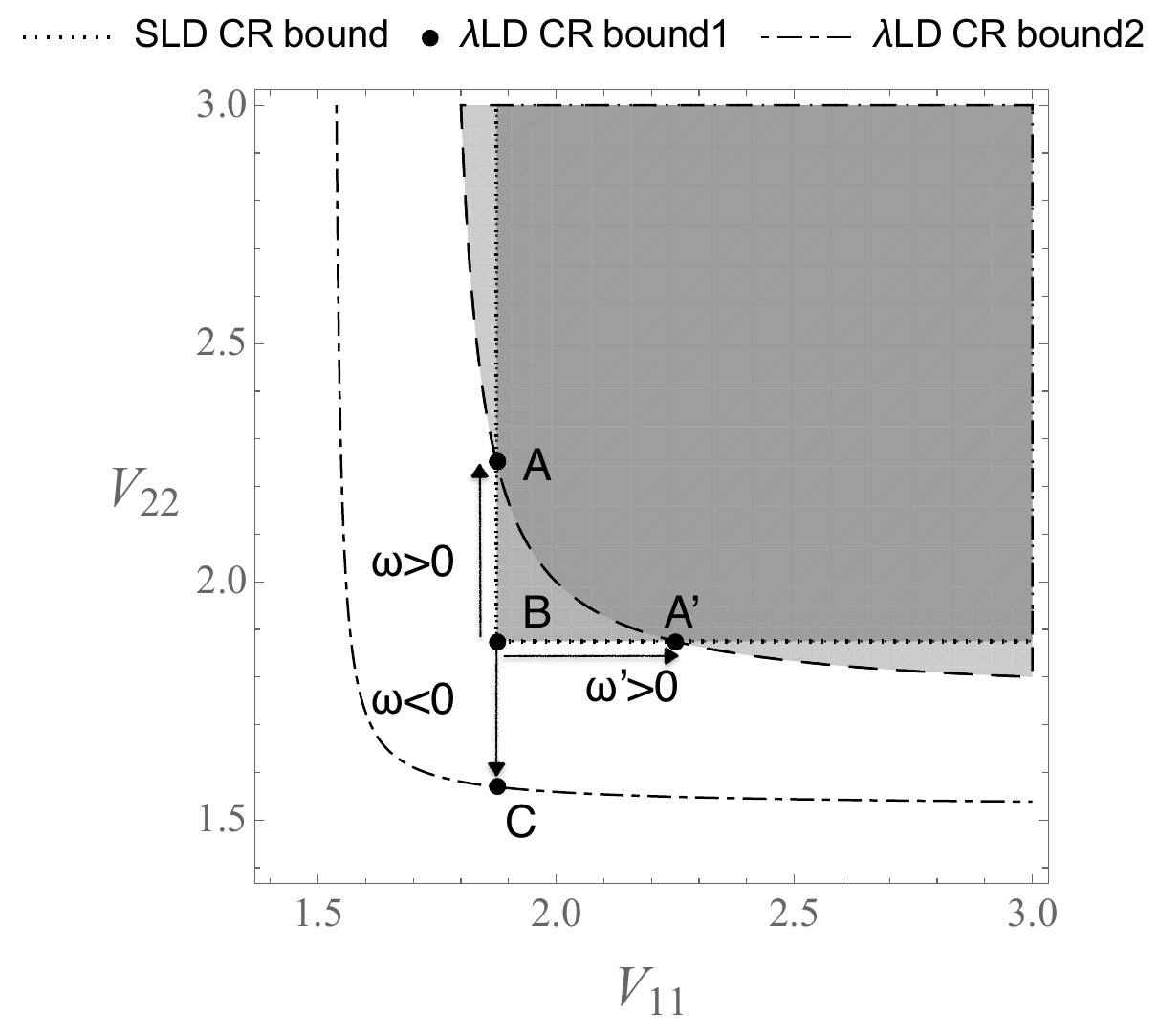}
\caption{A conceptual diagram showing cases with and without intersection points 
of the SLD and $\lambda$LD CR bounds. The dots show the points A, B, and C. 
The $\lambda$LD CR bounds 1 and 2 are examples of positive and negative $\omega$s, respectively. 
If $\omega$ is positive, $\omega'$ is also positive. 
When $\omega$ is positive, we can conclude that there exists a tradeoff relation 
because the SLD and the $\lambda$LD CR bounds intersect. 
 }
\label{fig2}
\end{center}
\end{figure}

Before moving to the definition of the indicator $\omega$, 
let us first discuss the logic of determining a tradeoff relation presented in Ref.~\cite{sf2} by
extending it to the $\lambda$LD Fisher information matrix. 
We will focus on two-parameter estimation, which is not necessarily a unitary shift model. 
We denote the mean-square-error (MSE) matrix for a locally unbiased estimator by $\sfV=[\sfV_{ij}]$. 
(We omit the dependence on the estimator and measurement since it is not important here.)
A graphical illustration of the indicator $\omega$ is given in Figure~\ref{fig2}. 
In this figure, we consider a special case where the SLD and the $\lambda$LD bounds are symmetric about 
$\sfV_{11}$ and $\sfV_{22}$. 
\begin{enumerate}
\item By a tradeoff relation, we mean a product type of inequality between 
the $(1,1)$ and $(2,2)$ components of the MSE matrix. 
That is $\sfV_{11}\sfV_{22}$ is bounded below by some constant. 

\item When one wishes to estimate one of the parameters, say $\theta_1$, 
it is known that we can perform a measurement that attains  $V_{11}=J_{S,11}^{-1}$. 
\footnote{
We remark that the ultimate limit to estimate 
$\theta_1$ is given by the $(1,1)$ component of the inverse of 
the SLD Fisher information matrix, $J_{S,11}^{-1}$, 
but not by 
$ (J_{\mathrm{S},11})^{-1}$ 
unless the SLD Fisher information matrix is diagonal. 
See Sec. 5 of Ref.~[31] for a detail account of the matter.} 
However, this optimal estimation strategy for $\theta_1$ cannot estimate $\theta_2$ in general, 
and hence $\sfV_{22}$ formally diverges. The same conclusion holds for estimating $\theta_2$. 
The SLD CR inequality $\sfV\ge J^{-1}_{\: \mathrm{S}}$ alone does not give 
any useful tradeoff relation since we only have 
\be
(\sfV_{11}-J^{-1}_{\: \mathrm{S} , 11})(\sfV_{22}-J^{-1}_{\: \mathrm{S} , 22})>0. 
\ee
The upper square region gives the allowed region for the diagonal components of the MSE 
in Fig.~\ref{fig2}. 

\item We next consider the $\lambda$LD CR inequality $\sfV\ge J^{-1}_{\: \lambda}$ in addition. 
Suppose that the $\lambda$LD CR bound is represented by the dashed curve in Fig.~\ref{fig2}. 
The SLD and the $\lambda$LD CR bounds have two intersection points in this case. 
(One of them is given by point A.) 
We then conclude that the tradeoff relation exists, since the allowed region 
is set by the combination of two CR bounds (the darker region in Fig.~\ref{fig2}). 

\item If the $\lambda$LD CR bound is represented by the dashed-dotted curve in Fig.~\ref{fig2}, on the other hand, two bounds do not have any intersection. 
In this case, we cannot conclude whether a tradeoff relation exists or not. 

\end{enumerate}

These motivate us to define the quantity $\omega$ and $\omega'$ 
that are shown in Figure~\ref{fig2}. 
The points ``A" and ``A'\," 
are the intersection points 
of the boundaries of the $\lambda$LD (dashed line) and SLD (dotted line) CR bounds. 
The indicator $\omega$ is defined by the $\sfV_{22}$ component of the point ``A" minus ``B"
which is equal to $J^{-1}_{\: \mathrm{S} , 22}$. 
When the $\lambda$LD and the SLD bounds do not intersect, on the other hand, 
the indicator $\omega$ is given by ``C" minus ``B," which is negative. 
The other quantity $\omega'$ in Figure~\ref{fig2} is 
also calculated in the same way.

As shown in Appendix~\ref{sec:omega}, the indicators $\omega$ and $\omega'$ are formally defined by
\begin{align} 
\omega = \frac{| \mathrm{Im} J^{-1}_{\: \lambda , 12}  |^2 - 
 (J^{-1}_{\: \mathrm{S} , 11}-J^{-1}_{\: \lambda , 11} )(J^{-1}_{\: \mathrm{S} , 22}-J^{-1}_{\: \lambda , 22})}
 {J^{-1}_{\: \mathrm{S} , 11}-J^{-1}_{\: \lambda , 11}}. \label{eq:omega2} \\
 \omega' = \frac{| \mathrm{Im} J^{-1}_{\: \lambda , 12}  |^2 - 
 (J^{-1}_{\: \mathrm{S} , 11}-J^{-1}_{\: \lambda , 11} )(J^{-1}_{\: \mathrm{S} , 22}-J^{-1}_{\: \lambda , 22})}
 {J^{-1}_{\: \mathrm{S} , 22}-J^{-1}_{\: \lambda , 22}}. \label{eq:omega3}
\end{align} 
Since in our model, $J^{-1}_{\: \mathrm{S} , 11}=J^{-1}_{\: \mathrm{S} , 22}$ 
and $J^{-1}_{\: \lambda , 11}=J^{-1}_{\: \lambda , 22}$ hold, 
we have $\omega=\omega'$. Hereafter, we use $\omega$ only.
As the quantum Fisher information matrices are a function of the spread $\kappa$ and the velocity of the observer $V$, the indicator also depends on them in addition to the choice of $\lambda\in(-1,1)$. 
From Eqs.~(\ref{ch7_eq:Jlambdainv}, \ref{eq:SLD_Fisher}, \ref{eq:omega2}), 
we have an explicit expression of the $\omega(\lambda, \kappa, {\v})$. 
\begin{align}
\omega(\lambda, \kappa, {\v}) 
=\frac{\kappa^2}{2 (1- \zeta^2)}
 \frac{\lambda ^2 (1-\zeta^2 )^2 - \xi ^2 [ \lambda ^2 (1-\zeta ^2)+ \zeta ^2 ]^2}
{ \xi ^2 [ \lambda ^2 (1- \zeta ^2)+  \zeta ^2 ] -(1-\zeta ^2)^2  }. 
\label{eq:explicit_omega2}
\end{align}
We set the allowed range of $\lambda$ 
as 0$<\lambda<$ 1 hereafter because $\omega(\lambda, \kappa, {\v})=\omega(-\lambda, \kappa, {\v})$ holds. 

\subsection{Existence of tradeoff relation}
\begin{figure}[t]
\begin{center}
\includegraphics[width=8.5cm]{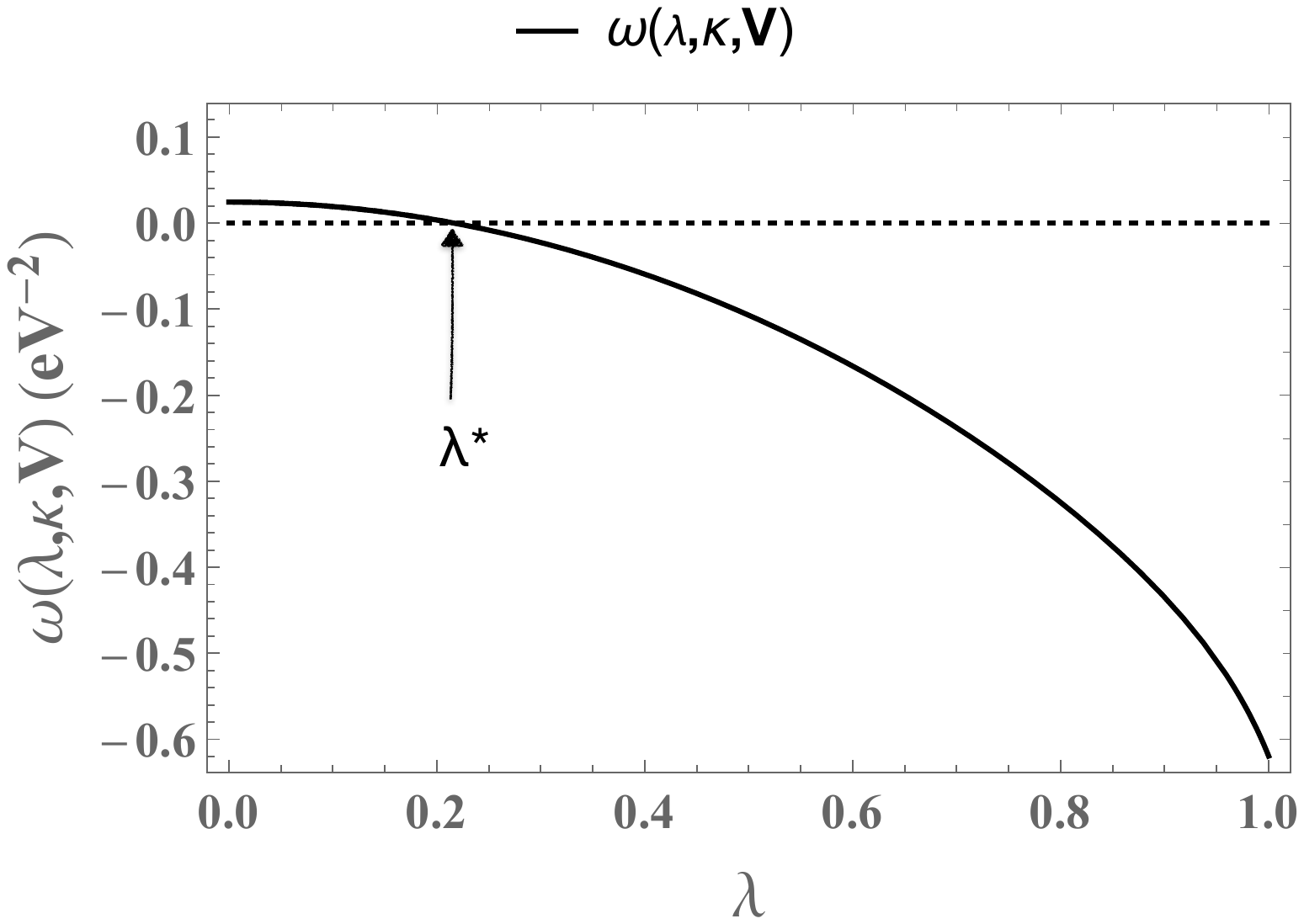}
\caption{Numerically calculated $\omega(\lambda, \kappa, {\v})$  
plotted as a function of $\lambda$ for $m \kappa=1$ and $V=1$ (relativistic limit). 
The dotted line shows the line for $\omega(\lambda, \kappa, \v)$=0. 
The $\lambda$ is dimensionless by definition. }
\label{ch7_fig_omega}
\end{center}
\end{figure}

We are ready to state the main result of this paper. 
\begin{quote}
\textit{For any spread of the wave function $\kappa> 0$ and in any moving frame, a nontrivial tradeoff relation exists, which is jointly specified by the SLD and $\lambda$LD CR bounds, as long as the $\lambda$ is smaller than the threshold value Eq.~\eqref{eq:kappa_ast}. In other words, any moving observer cannot estimate two parameters simultaneously, and the diagonal components of the MSE matrix can only take values in the darker region of Fig.~2.}
\end{quote}
This result can be proven by the following reasoning. 
First, the indicator $\omega(\lambda, \kappa, \v)$ is a continuous function of $\lambda$. 
Second, this is a monotonically decreasing function of $\lambda \in (0,1)$ for 
any given spread $\kappa > 0$ and for any given observer's velocity $\v \in (0, 1]$ (Appendix~\ref{sec:fuction_omega}). 
Third, this has different signs at two endpoints $\lambda=0,1$. 
In particular, the limit $\lambda \rightarrow0$ is always positive (Appendix~\ref{sec:limit0}), 
and the other limit $\lambda \rightarrow1$ is always negative (Appendix~\ref{sec:limit1}). 
Last, there always exists a unique solution 
$\lambda^\ast \in (0, 1)$ that satisfies $\omega(\lambda^\ast, \kappa, \v)=0$ for any $\kappa>0$ and $\v \in(0,1]$. 
Hence, we conclude that a tradeoff relation exists for our physical model 
in an arbitrary moving frame, no matter how slow the observer's velocity is. 
Figure~\ref{ch7_fig_omega} shows a numerically calculated $\omega(\lambda, \kappa, {\v})$  
plotted as a function of $\lambda$ for $m \kappa=1$ and $V=1$ (relativistic limit). 
This figure shows the above mathematical statements. 
The intersection point between the curve $\omega(\lambda, \kappa, {\v})$ and the line $\omega=0$ 
corresponds to the unique solution $\lambda^\ast$. 

We remark that there exists no tradeoff relation in the rest frame. 
This is because the parametric model in the rest frame is classical \cite{sf}. 
Indeed, in the present analysis $\zeta =0$ and $\xi=1$ hold in the limit $\v\to0$. 
This gives $\omega(\lambda, \kappa, \v=0) = -\kappa^2\lambda^2/2$ in the rest frame limit. 
Thus, we cannot conclude the existence of a tradeoff relation since this is negative. 
(See also discussion in Sec.~\ref{sec:model_rest}.) 
This means that we can estimate two parameters simultaneously in the rest frame. 
However, any observer in the moving frame cannot do this due to the existence of the tradeoff relation. 
We stress that the existence of the tradeoff relation is inevitable for {\it any} moving observer. 

Another interesting finding is that the tradeoff relation becomes the most significant 
in the limit of $\lambda \rightarrow 0$. 
This may not be expected, since the case $\lambda=0$ corresponds to the SLD Fisher information matrix. 
Thereby, one knows that it is impossible to detect a tradeoff relation. 
The resolution to this counter-intuitive result will be discussed in Sec.~\ref{sec:lambda_limit}. 

\subsection{Solution $\lambda^\ast$}
We next turn our attention to the unique solution $\lambda^\ast$ for the equation $\omega(\lambda, \kappa, {\v})=0$. 
We have shown that there exists a $\lambda^\ast \in (0,1)$ that is a unique solution of $\omega(\lambda^\ast, \kappa, {\v})=0$ 
for any $\kappa > 0 $ and any $\v \in (0,1]$. 
This is because the $\omega(\lambda, \kappa, {\v})$ is a monotonically decreasing function of $\lambda$ 
and because $\omega(0, \kappa, {\v})$ and $\omega(1, \kappa, {\v})$ are always positive and negative, respectively. 
Using the SLD CR bound and $\lambda$LD CR bound with any $\lambda \in (0, \lambda^\ast)$, we can conclude the existence of a tradeoff relation. 

\begin{figure}[t]
\begin{center}
\includegraphics[width=7cm]{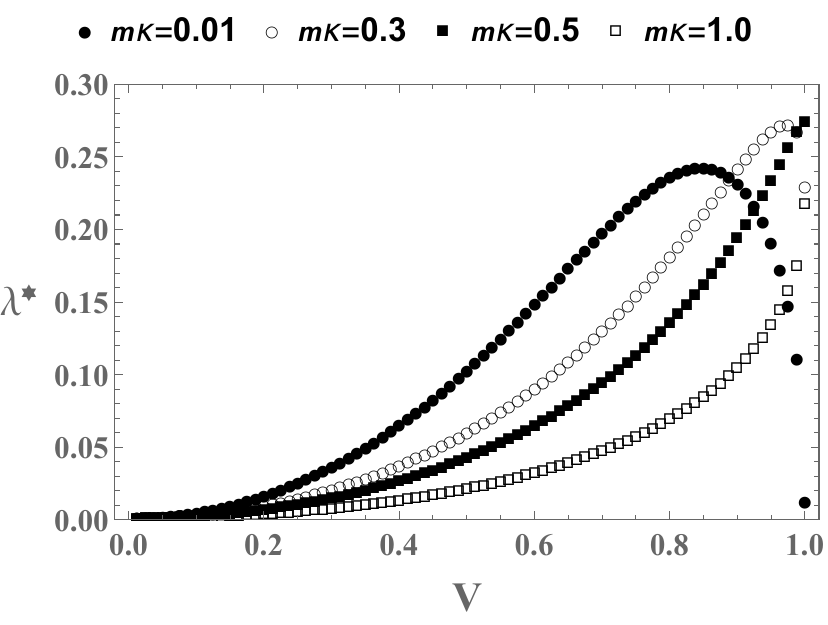}
\caption{Numerically calculated $\lambda^\ast$ plotted as a function of $\v$ for four different spread of the wave function. 
$\v$ and $\lambda^\ast$ are dimensionless in the natural units.
}
\label{fig4}
\end{center}
\end{figure}

In fact, it is possible to solve the equation $\omega(\lambda, \kappa, {\v})=0$ for $\lambda$ as 
a function of $\xi$ and $\zeta$. 
As shown in Appendix~\ref{sec:lambda_star_sol}, the unique solution $\lambda^\ast$ is expressed as
\begin{align}\label{eq:kappa_ast}
\lambda^\ast(\kappa,\v)
&=\frac{1}{2\xi} \left(1 - \sqrt{1 - \frac{4 \xi^2  \zeta ^2}{ 1-\zeta ^2 }} \, \right) .
\end{align} 
Figure~\ref{fig4} shows the result of the numerical calculation of $\lambda^\ast$ plotted as a function of $\v$ for four different values of the spread of the wave function. 
Let us analyze the properties of the solution $\lambda^\ast$. 
First, we see from Fig.~\ref{fig4} that $\lambda^\ast$ becomes smaller 
for slower velocities of the observer. In the rest frame limit $\v\to0$, 
we have $\lambda^\ast=0$ because $\zeta =0$ at $\v=0$. 
This is expected since no tradeoff relation exists in the rest frame.

Second, $\lambda^\ast$ is a convex upward function of $\v$ up to when $m \kappa$ is approximately less than 0.5. Otherwise, $\lambda^\ast$ is a monotonically increasing function of $\v$ for larger values of the wave function spread $\kappa$. 
The appearance of peaks for small values of $\kappa$ is not clear to us, 
in particular the existence of the velocity $\v$ that gives the maximum of $\lambda^\ast$. 
However, we point out that $\lambda$ is the parameter for the $\lambda$LD Fisher information matrix 
and is purely a mathematical quantity. 
The indicator $\omega$ is a more important quantity than $\lambda$, 
which is just a parameter because the $\omega$ directly indicates the strength of the tradeoff relation. 
We also discuss this point in the next remark regarding the relativistic limit.

Last, Fig.~\ref{fig4} shows that $\lambda^\ast$ is close to zero at the relativistic limit $\v=1$ 
when $\kappa \ll 1$. 
The following relation supports this sudden drop. 
\be
\lim_{\kappa \to 0} \lambda^\ast (\kappa, 1)=0.  \label{eq:lambda_star}
\ee
It also explains the appearance of peaks for small $\kappa$. 

A proof for Eq.~\eqref{eq:lambda_star} is given as follows. 
At the relativistic limit $\v=1$ and in the limit of $\kappa \rightarrow 0$, we have 
$\zeta \rightarrow \sqrt{2\pi}/4$ and $\xi \rightarrow 0$ (see Appendix~\ref{sec:rel_limit}). 
We apply the Taylor expansion for small $\xi$ for $\lambda^\ast$ as follows.
\be
\lambda^\ast
= \frac{1}{2\xi}\sum_{n=1}^\infty  \frac{(2n-3)!!}{n! 2^{n}} \left( \frac{4 \xi^2 \zeta ^2}{ 1-\zeta ^2 } \, \right)^n,
\ee
where
\be
(2n-3)!!=(2n-3) (2n-5) \cdots 3 \cdot 1, \quad (-1)!! = 1.
\ee
The first order in $\xi$ is thus given by $\zeta^2\xi/(1-\zeta^2)$. 
Therefore, at $\v=1$, Eq.~\eqref{eq:lambda_star} holds. 

\section{Discussion}
\subsection{Comparison to the rest frame}\label{sec:model_rest}
The parametric model in the rest frame $\mathcal{M}_{\rm rest}$, Eq.~\eqref{eq:model_rest} is 
classical. This is because the reference state is a Gaussian state 
which is a product of two Gaussian functions of $p^1$ and $p^2$. 
Furthermore, the generators of the unitary model, $\hat{p}^1$ and $\hat{p}^2$ commute, 
the best estimate is obtained by the position measurement of each of $x$ and $y$ independently in the rest frame. 
By this optimal measurement, one can show that the MSE matrix is equal to the inverse of 
the SLD Fisher information matrix. 
 
On the other hand, 
the model in the moving frame, $\mathcal{M}^{\Lambda}$, Eq.~\eqref{eq:model_rel} 
would change from a classical to a nonclassical, nontrivial model since the wave function 
after the Lorentz boost changes to a more complicated form due to the observer's motion. 
This change results from taking the partial trace of the entangled state generated by the Wigner rotation. 
Mathematically, this corresponds to the action of a completely-positive and trace-preserving (CP-TP) map 
on the initial classical state. Here, the spin degree freedom plays the role of an environment in 
the standard terminology of quantum information theory. 
Thus, the moving observer faces estimating a `noisy' state caused by this CP-TP map. 

While this is a mathematical fact, it is highly nontrivial that an application of such a CP-TP map on a `classical' state gives rise to the tradeoff relation. 
This tradeoff relation prohibits any moving observer from estimating the two parameters simultaneously. 
The only possibility to avoid incompatibility is to measure not only the position of the particle but also 
the spin. 

Lastly, let us examine the rest frame limit. 
We excluded the case $\v=0$ in the above analysis, corresponding to a nonmoving observer. 
By definition, two integrals Eqs.~\eqref{eq:zeta} and \eqref{eq:xi} are evaluated analytically as $\zeta =0$ and $\xi=1$ hold in the limit $\v\to0$. 
This results in $\omega(\lambda, \kappa, \v=0) = -\kappa^2\lambda^2/2<0$ as mentioned earlier. 
Thereby, we cannot conclude the existence of a tradeoff relation. 
The unique solution $\lambda^\ast$ \eqref{eq:kappa_ast} becomes zero in this limit. 
The other explanation is that the $\lambda$LD Fisher information matrix 
becomes proportional to the SLD Fisher information matrix in this limit. 
Thus, there is no way to evaluate a tradeoff relation for any $\lambda$. 
\subsection{Strength of tradeoff relation} \label{sec:lambda_limit}
As explained in Fig~\ref{fig2}, when the indicator $\omega$ is positive, 
we can conclude that a tradeoff relation exists. 
Indeed we have shown that this is true for any given spread of the wave function, $\kappa > 0$, and any given velocity of the observer $0<\v \leq 1$. 
The tradeoff relation is most substantial in the limit of $\lambda \rightarrow 0$ while at $\lambda=0$, by definition, the $\lambda$LD Fisher information matrix is equal to the SLD Fisher information matrix. 
As we stressed before, the SLD Fisher information alone cannot determine the existence of a tradeoff relation. 
This counter-intuitive result is explained as follows. 

The key quantity in our analysis is the indicator $\omega$, Eq.~\eqref{eq:omega2}. 
This quantity is defined by the ratio between the differences between two quantum Fisher information matrices. 
Thus, this ratio can be finite even when each difference becomes zero. 
To see this point more clearly, we rewrite the numerator and denominator of Eq.~\eqref{eq:omega2} as  
\begin{align}
-&\det (J^{-1}_{\: \mathrm{S}}-J^{-1}_{\: \lambda}),\nonumber\\
&\frac12\tr (J^{-1}_{\: \mathrm{S}}-J^{-1}_{\: \lambda}),\label{eq:difference}
\end{align}
respectively. 
These alternative expressions are true since both quantum Fisher information matrices have symmetry 
and the off diagonal components of the $J^{-1}_{\: \lambda}$ are purely imaginary. 
It is true that in the limit $\lambda\to0$, both terms in Eq. \eqref{eq:difference}, 
$\tr J^{-1}_{\: \mathrm{S}}$ 
and 
$\tr J^{-1}_{\: \lambda}$ become 0. 
However, their ratio does have a meaningful limit. 
In the present case, both terms are proportional to $\lambda^2$; hence, we have the following well-defined limit. 
\begin{equation}\label{eq:lambda_max}
\lim_{\lambda\to0} \omega(\lambda, \kappa, \v)=
\frac{\kappa^2}{2 (1- \zeta^2)}
 \frac{  \xi ^2 \zeta ^2 }
{ (1-\zeta ^2)^2-\xi^2\zeta^2  } .
\end{equation}
\subsection{Dependence on spread of wave function $\kappa$}
Next, we discuss the role of the spread of the wave function in the rest frame. 
Intuitively, the smaller $\kappa$ is, the better the estimate is. 
This is because a wave function with a sharper peak enables us to 
estimate the unitary-shift more accurately. 
Indeed, this is so in the rest frame, in which the inverse of the SLD Fisher information matrix 
is given by $J^{-1}_{\: \mathrm{S}}=\kappa^2/2 I$ with $I$ the identity matrix. 
Thus, the infinitely sharp wave function (Dirac delta function) gives estimation precision with infinite accuracy.  
However, we show that this small value of $\kappa$ does not necessarily give a better estimate in the moving frame. 
In fact, using a sharper wave function can face a more significant tradeoff relation. 

In passing we note that a similar conclusion was obtained in a different problem when 
estimating gravitational redshift based on the quantum field theory in curved-space time \cite{bruschi}. 
In particular, this reference showed that sharper peaks are always better. 
The limit of spread $\kappa \rightarrow 0$ is also properly discussed. 
However, the current paper only concerns a finite width for simplicity. 
\begin{figure}[t]
\begin{center}
\includegraphics[width=8cm]{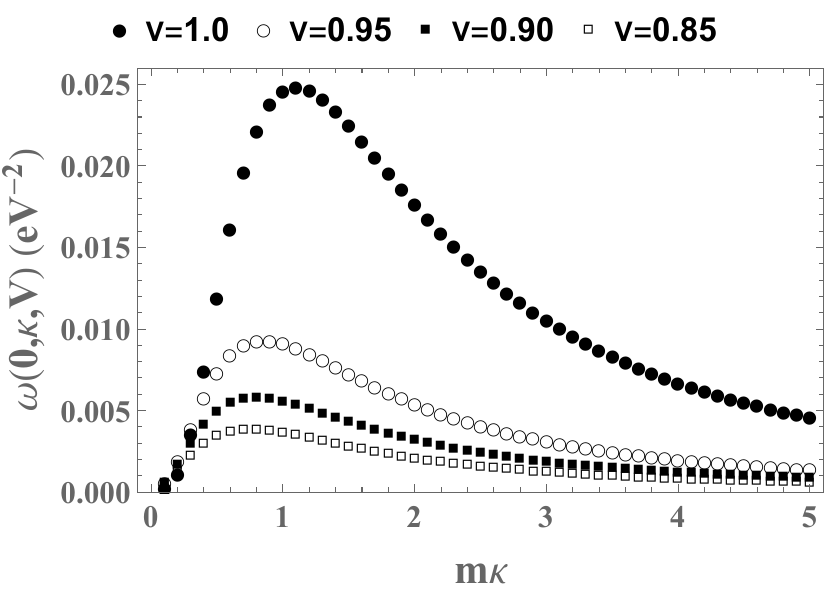}
\caption{Numerically calculated $\omega(0, \kappa, {\v})$  
plotted as a function of $m \kappa$ at $\v$=0.85, 0.9, 0.95, and 1.}
\label{fig5}
\end{center}
\end{figure} 
In Fig.~\ref{fig5}, we plot the maximum strength of the tradeoff relation $\omega(\lambda=0,\kappa,\v)$, Eq.~\eqref{eq:lambda_max},  
as a function of $m\kappa$ for four different velocities of a moving observer. 
From Fig.~\ref{fig5}, we see that $\omega(\lambda=0,\kappa,\v)$ has a one peak at some value of the spread $\kappa^\ast$. 
The peak positions depend on the velocity $\v$. 
We observe that the dependence on the velocity is monotonically increasing, 
that is, the faster the observer moves, the larger $\kappa^\ast$ is. 
This results in more significant tradeoff relations. 

It is counter-intuitive that the tradeoff relation can be the most significant 
with a specific spread of the wave function in the rest frame.  
We expect that this phenomenon has something to do with the separation distance between the peaks of
the spin-down and the spin-up wave functions.  
We have no physically clear explanation for this result at this moment. 
\section{Conclusion}
We investigated the tradeoff relation to estimate the unitary-shift parameters for a moving observer. 
The model considered in this paper was studied, 
but we were unable to discuss a tradeoff relation in our previous study~\cite{sf}. 
This was because only the SLD CR bound was analyzed previously. 
In this paper, we extend the idea of combining different types of quantum Fisher information matrices, 
which was proposed in Ref.~\cite{sf2}. 
This was done by combining the standard SLD CR bound and the $\lambda$LD CR bound 
to evaluate the proposed indicator $\omega$, Eq.~\eqref{eq:omega2}. 

By the proposed method using the indicator $\omega$, we proved a nontrivial tradeoff relation 
for any moving observer. Further, this existence of the tradeoff relation was 
analyzed in detail, in particular its dependence on the physical parameters such as 
the spread of the wave function and the velocity of the observer. 

Before closing the paper, we wish to emphasize the versatility of the proposed indicator in the sense that 
it can be applied to any parametric model to discuss a tradeoff relation. 
This method is simple since we only need to calculate the $\lambda$LD Fisher information matrix. 
We plan to extend it to a multiple-parameter model in future work.  
Applications to other physical models are also an interesting topic.

\section*{Acknowledgment}
We would like to express our gratitude for invaluable discussions 
to Prof. Yoko Miyamoto and Prof. Hiroshi Nagaoka, of The University of Electro-Communications. 
We also would like to thank anonymous reviewers for careful reading, constructive comments, 
and valuable suggestions. 
The work is partly supported by JSPS KAKENHI Grant Number JP21K11749. 
%
\appendix
\section{Wave function in the moving frame} \label{sec:wavefunction}
We briefly summarize the Wigner rotation to derive the state vector in a moving frame. 
For more detailed derivation, see~\cite{sf}.

\subsection{Lorentz transformation}
Let $\Lambda$ be a Lorentz transformation from the rest frame to the moving frame,  
\begin{align}
\Lambda&=
\left(
\begin{array}{cccc}
 \cosh \chi  & 0 & 0 & -\sinh \chi  \\
 0 & 1 & 0 & 0 \\
 0 & 0 & 1 & 0 \\
 -\sinh \chi  & 0 & 0 & \cosh \chi  \\
\end{array}
\right),  \\
\cosh \chi &= \frac{1}{\sqrt{1-V^2}}, \quad \sinh \chi = \frac{V}{\sqrt{1-V^2}}. \label{eq:Lambda}
\end{align}
where $V$ is the velocity of the observer, 
We assume that the observer's motion is along the $z$ axis. 
The four-momentum of the particle is transformed as 
\be
\Lambda p = ( (\Lambda p)^0, (\Lambda p)^1, (\Lambda p)^2, (\Lambda p)^3)  
\equiv ( (\Lambda p)^0, \overrightarrow{\Lambda p}).
\ee
Here we define the spatial part of the four-momentum by$ \overrightarrow{\Lambda p}$. 

By the above Lorentz transformation $\overrightarrow{\Lambda p}$ is 
\begin{align}
\overrightarrow{\Lambda p}
&= \left(\sum_{\mu=0}^3 \Lambda^1_{\: \mu} p^\mu,\, \sum_{\mu=0}^3 
\Lambda^2_{\: \mu} p^\mu,\, \sum_{\mu=0}^3 \Lambda^3_{\: \mu} p^\mu\right) \nonumber \\
&=(p^1, p^2, -p^0 \sinh \chi), 
\end{align}
where $p^0=\sqrt{m^2+ \p^2}$ and $\p^2=( p^1)^2+(p^2)^2$. 

\subsection{State in the moving frame: Wigner rotation} \label{sec:Wigner}
The state vector in the moving frame is related to that in the rest frame by a unitary 
transformation $U(\Lambda)$ ~\cite{halpern,weinberg}. 
For a spin-1/2 particle with a mass $m$, this relation is given as
\be
U(\Lambda) \ket{p, \sigma}
= \sqrt{\frac{(\Lambda p)^0}{p^0}} \sum_{\sigma^\prime=\downarrow, \uparrow} D^{(\frac{1}{2})}_{\sigma^\prime, \sigma} (W(\Lambda, p)) \ket{\Lambda p, \sigma},
\label{eq:Ulambdap} 
\ee
where $D^{(\frac{1}{2})}_{\sigma^\prime, \sigma} (W(\Lambda, p)) $ is the spin-1/2 representation of 
a three-dimensional rotational group that is determined by $W(\Lambda,p)=L^{-1}(\Lambda p) \Lambda L(p)$ 
with $L(p)$ defined below. 
In the Wigner rotation description of the Lorentz transformation, 
the essential part is to use the spatial part of $W(\Lambda,p)$. 
This then gives a rotation on the Pauli spin by the standard correspondence. (see for example  Ref.~\cite{sakurai}) 
We choose $L(p) =[ L^i_{\: j}(p) ]$ as the one given in~\cite{weinberg}. 
\begin{align}
L^i_{\: j} (p) &=\delta_{i j} + \frac{(\sqrt{m^2+ \p^{\, 2}} -m) p^i p^j}{ m \p^{\, 2}}, \nonumber \\
L^i_{\: 0} (p) &=\frac{p^i}{m}, \nonumber \\
L^0_{\: 0} (p) &=\frac{\sqrt{m^2+ \p^{\, 2}}}{m}. \nonumber
\end{align}
By setting $\vec{p}=(p^1, p^2, 0)$ for our model, we obtain $W(\Lambda,p)$ as  
\begin{align}
[W(\Lambda, p)]^0_{\: \: 0}&=1, \nonumber \\
[W(\Lambda, p)]^1_{\: \: 0}&=[W(\Lambda, p)]^0_{\: \: 1}=0, \nonumber \\
[W(\Lambda, p)]^2_{\: \: 0}&=[W(\Lambda, p)]^0_{\: \: 2}=0, \nonumber \\
[W(\Lambda, p)]^3_{\: \: 0}&=[W(\Lambda, p)]^0_{\: \: 3}=0, \nonumber
\end{align}
\begin{widetext}
\be
[W(\Lambda, p)]^1_{\: \: 1}=[W(\Lambda, p)]^2_{\: \: 2}
=\frac{ p^0 [m (p^1)^2 + p^0 (p^2)^2] \sinh ^2\chi + \p^2 [ (p^1)^2 \cosh \chi+(p^2)^2]}
{\p^2 \left[(p^0)^2 \sinh ^2\chi+\p^2 \right]},  \nonumber
\ee
\end{widetext}
\begin{align}
[W(\Lambda, p)]^2_{\: \: 1}&=[W(\Lambda, p)]^1_{\: \: 2} 
 = - \frac{p^1 p^2 ( \cosh \chi  -1) (p^0-m)}{\p^2 (p^0 \cosh \chi + m)},  \nonumber \\
[W(\Lambda, p)]^3_{\: \: 1}&=-[W(\Lambda, p)]^1_{\: \: 3}
=- \frac{p^1  \sinh \chi}{p^0 \cosh \chi + m}, \nonumber \\
[W(\Lambda, p)]^3_{\: \: 2}&=-[W(\Lambda, p)]^2_{\: \: 3} 
= - \frac{p^2  \sinh \chi}{p^0 \cosh \chi + m},  \nonumber \\
[W(\Lambda, p)]^3_{\: \: 3}
&=\frac{p^0 + m \cosh \chi}{m + p^0 \cosh \chi}. \nonumber
\end{align}

We next define a $3\times3$ real matrix $[R(\Lambda, p)]_{j k}$ by the spatial part of $W(\Lambda, p)$ as 
\be
[R(\Lambda, p)]_{jk}= [W(\Lambda, p)]^j_{\: \: k} \, , \quad (j,k=1,2,3). \nonumber
\ee
We now convert this three-dimensional rotation matrix into the spin-1/2 representation 
to get the desired Wigner rotation. 
This can be done by decomposing $R(\Lambda, p)$ into three Euler angles as 
(In fact, we only need two angles in our example. A direct calculation can show this.)
\be
R(\Lambda, p)
= R_z(-\phi) R_y( \alpha) R_z( \phi),  \label{eq:EulerRotation}
\ee
where
\begin{align}
R_y( \alpha)&=\begin{pmatrix}
\cos \alpha & 0 & -\sin \alpha \\
0 & 1 & 0 \\
\sin \alpha & 0 & \cos \alpha 
\end{pmatrix}, \nonumber \\
R_3( \phi)&=\begin{pmatrix}
\cos \phi & -\sin \phi & 0 \\
\sin \phi & \cos \phi & 0 \\
0 & 0 & 1 
\end{pmatrix}. \nonumber
\end{align}
We finally obtain 
\begin{align}
D^{(\frac{1}{2})}(W(\Lambda,p))&= \e^{ \I  \frac{\phi}{2}\sigma_3}  \e^{- \I  \frac{\alpha}{2}\sigma_2}  \e^{ -\I  \frac{\phi}{2}\sigma_3}\\
&= \begin{pmatrix}
\cos \frac{\alpha}{2}  & - \e^{\I \phi} \sin \frac{\alpha}{2} \\
 \e^{-\I \phi} \sin \frac{\alpha}{2} & \cos \frac{\alpha}{2} 
\end{pmatrix}. \nonumber
\end{align}
This $D^{(\frac{1}{2})}(W(\Lambda,p))$ gives Eqs.~\eqref{eq:Psilambda},~\eqref{eq:Psi_pi},~\eqref{eq:F1}, and~\eqref{eq:F2} after straightforward calculations.
\subsection{Wave function after boost in the $x$ representation} \label{sec:peak_position}
%
In this section, we prove that the spin-up probability density 
has rotational symmetry around the peak of the original Gaussian wave packet. 
To see this symmetry, we can set $\theta=0$. 
The state vector after the Lorentz boost $\ket{\psi^\Lambda_{\: \sigma}(\theta=0)}$ is expressed as 
\begin{align*}
 \ket{\psi^\Lambda_{\: \sigma}(\theta=0)}&= \int d^3p \sqrt{\frac{(\Lambda p)^0}{p^0}} F_{\theta=0, \, \sigma}({p}^1, {p}^2) \delta(p^3) 
 \ket{\overrightarrow{\Lambda p}}.
 \end{align*}
Then, its Fourier transform is 
\begin{align*}
 \braket{x | \psi^\Lambda_{\: \uparrow}(\theta=0)}&= \int d^3p \sqrt{\frac{(\Lambda p)^0}{p^0}} F_{\theta=0, \, \uparrow}({p}^1, {p}^2) \delta(p^3) 
 \braket{x | \overrightarrow{\Lambda p}} \nonumber \\
 &=-\sqrt{\frac{2}{1-\xi}} \frac{\kappa}{(2 \pi)^2} \sqrt{\cosh \chi} \\
    & \quad \times \int dp^1 dp^2 \e^{- \frac{1}{2} \kappa^2 [(p^1)^2+(p^2)^2]+ \I \phi(p^1, \, p^2)} \\
    & \quad \times \sin \frac{\alpha(\p)}{2} \e^{-\I p^1 x^1 -\I p^2 x^2 -\I \sqrt{(p^1)^2+(p^2)^2 +m^2} \sinh \chi x^3}. \nonumber \\
 \end{align*}
 To execute the integration, we switch to the polar coordinate ($p,\Phi$) from the $(p^1,p^2)$ coordinate. Then, 
 the integration over $\Phi$ is expressed by
 \be
\int_0^{2\pi} d\Phi \e^{\I\Phi} \e^{-\I p^1 x^1 - \I p^2 x^2}
= \int_0^{2\pi} d\Phi \e^{\I\Phi} \e^{-\I p (x^1\cos \Phi + x^2\sin \Phi)}. 
 \label{eq: wave_funtion_angle}
 \ee
 We use 
 \begin{align}
 p^1=p \cos \Phi, \, 
 p^2=p \sin \Phi. 
 \end{align}
 Let us express $x^1$ and $x^2$ as follows.
 \begin{align}
 x^1=r \cos \delta, \,
 x^2=r \sin \delta,  
 \end{align}
 where $r=\sqrt{(x^1)^2+(x^2)^2}$. 
 By a straightforward calculation, Eq.~\eqref{eq: wave_funtion_angle} is expressed as follows.
 \begin{align}
\int_0^{2\pi} d\Phi \e^{\I\Phi} \e^{-\I p^1 x^1 - \I p^2 x^2}
&=\int_0^{2 \pi} d\Phi \e^{-\I p r \cos( \Phi - \delta)+ \I \Phi} \nonumber \\
&=2 \e^{\I \delta} \int_0^{ \pi} d\Phi  \cos \Phi \e^{-\I p r \cos \Phi} .\nonumber 
\end{align}
Using the Hasen-Bessel formula, 
\be
J_n(x)= \frac{1}{\pi \I^n} \int_0^{ \pi} d \theta \cos n \theta \, \e^{ \I x \cos \theta}, 
\ee
where $J_n(x)$ is a Bessel function of the $n$th kind, we have
\be
\int_0^{\pi} d\Phi \e^{\I\Phi} \e^{-\I p^1 x^1 - \I p^2 x^2}
= - 2 \pi \I \e^{ \I \delta} J_1(\p r). 
\ee
Substituting this expression into the Fourier transform gives
\begin{align*}
 \braket{x | \psi^\Lambda_{\: \uparrow}(\theta=0)}
 &=\I \sqrt{\frac{2}{1-\xi}} \frac{\kappa}{2 \pi} \sqrt{\cosh \chi} \e^{ \I \delta}  \\
    & \quad \times \int_0^\infty dp p \e^{- \frac{1}{2}\kappa^2 p^2} J_1(p r)
   \sin \frac{\alpha(p)}{2} \e^{ -\I p^0 x^3\sinh \chi }. \nonumber \\
 \end{align*} 
Therefore, we see that $|  \braket{x | \psi^\Lambda_{\: \uparrow}(\theta=0)} |^2$ has an axial symmetry 
around the $z$ axis in the $x$-$y$ plane because it does not depend on the angle $\delta$.
\section{$\lambda$LD for an $n$-parameter rank deficient model} \label{ch7_sec:lambdaLD_no_full_rank}
Consider the general $p$-parameter model which is not necessarily full-rank:
\be
\mathcal{M} =\{\rho_\theta | \theta=(\theta_1, \theta_2,\cdots,\theta_p) \in \Theta \},
\ee
where rank $\rho_\theta = r \leq d =$ dim $\mathcal{H}$ for all $\theta \in \Theta$. 
We derive the $\lambda$LD Fisher information matrix at $\theta$. 
In the following, we drop $\theta$ since we are only concerned with the fixed $\theta$. 
To proceed with our calculation, we introduce the following index convention. 
\begin{quote}
 \begin{itemize}
  \item $\alpha, \beta, \gamma, \cdots$ for $\{1,2,\cdots,d \}$: All indices
  \item $i, j, k, \cdots$ for $\{1,2,\cdots,r \}$: Support of $\rho_\theta$
  \item $a, b, c, \cdots$ for $\{r+1,\cdots,d\}$: Kernel of $\rho_\theta$
  \item $m,n,\cdots$ for the parameter index
 \end{itemize}
\end{quote}

Consider the eigenvalue decomposition of the state $\rho$ as
\be
\rho = \sum_{i=1}^r \rho_i \ket{e_i} \bra{e_i}. \label{eq:rho_expansion}
\ee

If we use $\rho_a=0$ (zero eigenvalue for $a=r+1,\ldots,d$) for the kernel space of $\rho$, we can also write
\be
\rho = \sum_{\alpha=1}^d \rho_\alpha \ket{e_\alpha} \bra{e_\alpha},
\ee
by appropriate orthonormal vectors $\bra{e_a}$ for the kernel space. 
The $\lambda$LD for $\theta_n$ is defined by a solution to
\be
\partial_n \rho = \frac{1+\lambda}{2} \rho L_n +  \frac{1-\lambda}{2} L_n \rho, 
\ee
where 
\be
\partial_n= \frac{\partial}{\partial \theta_n}.
\ee
We expand $L_n$ in the $\ket{e_\alpha}$ basis as
\be
L_n= \sum_{\alpha,\beta=1}^d l^{(n)}_{\alpha,\beta} \ket{e_\alpha} \bra{e_\beta}.
\ee
The coefficients $ l^{(n)}_{\alpha,\beta}$ are determined by
\be
\braket{e_\alpha | \partial_n \rho | e_\beta}
= \left[ \frac{1+\lambda}{2} \rho_\alpha + \frac{1-\lambda}{2} \rho_\beta \right]  l^{(n)}_{\alpha,\beta}.
\ee
This equation determines  $l^{(n)}_{\alpha,\beta}$ for $\alpha, \beta \notin \{ r+1, \cdots, d \}$ only.
\be
  \lambda_{\alpha, \beta}=
  \begin{cases}
    \frac{1+\lambda}{2} \rho_\alpha + \frac{1-\lambda}{2} \rho_\beta & \text{for $\alpha, \beta \notin \{ r+1, \cdots, d \}$,} \\[2ex]
    \text{indetermined}       & \text{otherwise.}
  \end{cases}
\ee
For convenience, we denote $\lambda^{\pm}_i =  \frac{1 \pm \lambda}{2} \rho_i$, then we have 
\begin{align}
\lambda_{i, a} = \lambda^{+}_{\: i}, \\
\lambda_{a, i} = \lambda^{-}_{\: i}.
\end{align}

The $\lambda$LD $L_n$ is obtained as
\be
L_n= \sum_{\alpha,\beta} \nolimits'  \lambda^{-1}_{\alpha,\beta} \ket{e_\alpha} \braket{e_\alpha | \partial_n \rho | e_\beta} \bra{e_\beta}, 
\ee
where the prime indicates summing over $\alpha, \beta \notin \{ r+1, \cdots, d \}$. 
By using the projectors $P_i=\ket{e_i}\bra{e_i}$ ($i=1,2,\ldots,r$), we can express
\begin{align}
L_n
&= \sum_{i=1}^r (\lambda^{+}_{\:i})^{\:-1} P_i \partial_n \rho + \sum_{j=1}^r (\lambda^{-}_{\:j})^{-1}  \partial_n \rho P_j \nonumber \\
&+ \sum_{i,j=1}^r \left[( \lambda_{i,j})^{-1} -(\lambda^{+}_{\:i})^{-1}-(\lambda^{-}_{\:j})^{-1} \right] P_i \partial_n \rho P_j .
\end{align}
We obtain an alternative expression by substituting Eq.~\eqref{eq:rho_expansion}.
\begin{align}
L_n
&= \sum_{i=1}^r \frac{\partial_n \rho_i}{\rho_i} + \sum_{i=1}^r (\lambda^{+}_{\:i})^{\:-1} \ket{e_i} \bra{\partial_n e_i}
+\sum_{i=1}^r (\lambda^{-}_{\:i})^{\:-1} \ket{\partial_n e_i} \bra{e_i} \nonumber \\
&+ \sum_{i,j=1}^r \left[( \lambda_{i,j})^{-1} -(\lambda^{+}_{\:i})^{-1} \right] \rho_i \braket{\partial_n e_i | e_j} \ket{e_i} \bra{e_j} 
\nonumber \\
&+ \sum_{i,j=1}^r \left[( \lambda_{i,j})^{-1} -(\lambda^{-}_{\:i})^{-1} \right] \rho_j \braket{ e_i | \partial_n e_j} \ket{e_i} \bra{e_j} .
\end{align}
By its definition Eq.~\eqref{eq:def_lambdaLD_FIM}, 
the $(m,\,n)$ component of the $\lambda$LD Fisher information matrix 
$J_{\lambda,mn}$ is calculated by
\be
J_{\lambda,mn}= \tr (\partial_n \rho L^\dagger_m).
\ee
The final expression for $J_{\lambda,mn}$ is
\begin{align}
J_{\lambda,mn}
&= \sum_{i=1}^r (\lambda^{+}_{\:i})^{\:-1} \braket{e_i | \partial_n \rho  \partial_m \rho | e_i}
+\sum_{i=1}^r (\lambda^{-}_{\:i})^{\:-1} \braket{e_i | \partial_m \rho  \partial_n \rho | e_i} \nonumber \\
&+ \sum_{i,j=1}^r \left[( \lambda_{i,j})^{-1} -(\lambda^{+}_{\:i})^{-1}-(\lambda^{-}_{\:j})^{-1} \right] 
\braket{e_i | \partial_n \rho | e_j} \braket{e_j | \partial_m \rho | e_i}. 
\end{align}
With further calculation, we have
\begin{align}
J_{\lambda,mn}
&= \sum_{i=1}^r \frac{\partial_n \rho_i \partial_m \rho_i}{\rho_i} \nonumber \\
&+\sum_{i=1}^r (\lambda^{+}_{\:i})^{\:-1}  (\rho_i)^2 \braket{\partial_n e_i | \partial_m e_i}
+\sum_{i=1}^r (\lambda^{-}_{\:i})^{\:-1}  (\rho_i)^2 \braket{\partial_m e_i | \partial_n e_i} \nonumber \\
&+ \sum_{i,j=1}^r \left[( \lambda_{i,j})^{-1}(\rho_i-\rho_j)^2 
-(\lambda^{+}_{\:i})^{-1} (\rho_i)^2-(\lambda^{-}_{\:j})^{-1} (\rho_j)^2 \right] \nonumber \\
& \quad \times \braket{e_i  | \partial_n e_j} \braket{\partial_m e_j  | e_i}. 
\label{app_eq:Jlambda_formula}
\end{align}

\section{Cram\'er-Rao bound determined by SLD and $\lambda$LD CR bounds} \label{sec:omega}
In this section, we explain the basic idea of the proposed method to determine the existence of a tradeoff relation. 
Then, we define the indicators $\omega$ and $\omega'$, 
Eqs.~(\ref{eq:omega}, \ref{eq:omega2}) of witnessing the existence of a tradeoff relation. 
For simplicity, we consider a special case of two-parameter models whose quantum Fisher information matrix is 
symmetric with respect to the index as an example. 
The generalization to a nonsymmetric case is straightforward. 

\begin{figure}[t]
\begin{center}
\includegraphics[width=8cm]{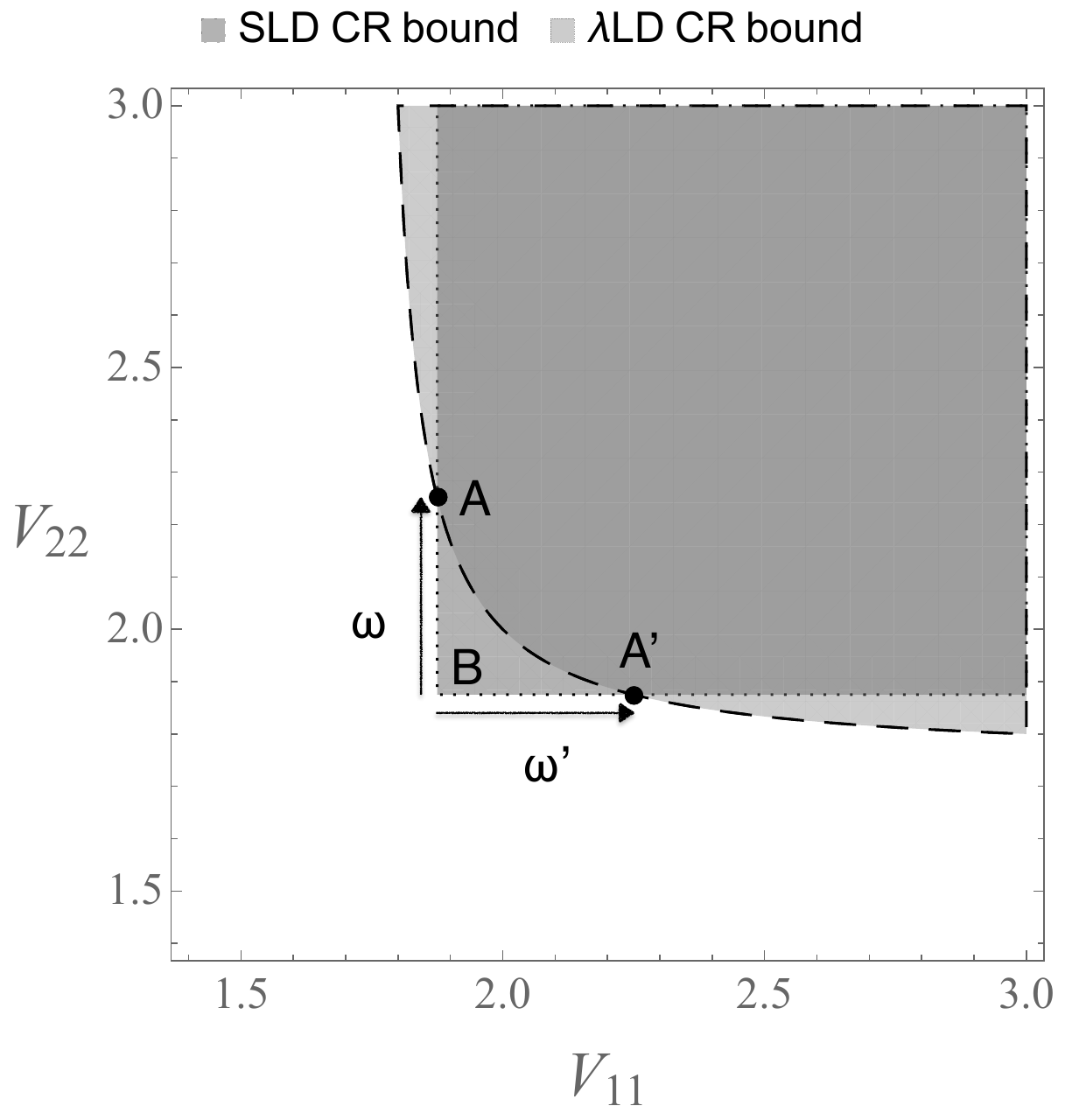}
\caption{The case of $\omega >0$ and $\omega'>0$. 
The intersection of the SLD and the $\lambda$ bounds 
(darker gray) is not the same as the SLD or $\lambda$LD bounds.}
\label{fig_app1}
\end{center}
\end{figure}

\begin{figure}[t]
\begin{center}
\includegraphics[width=8cm]{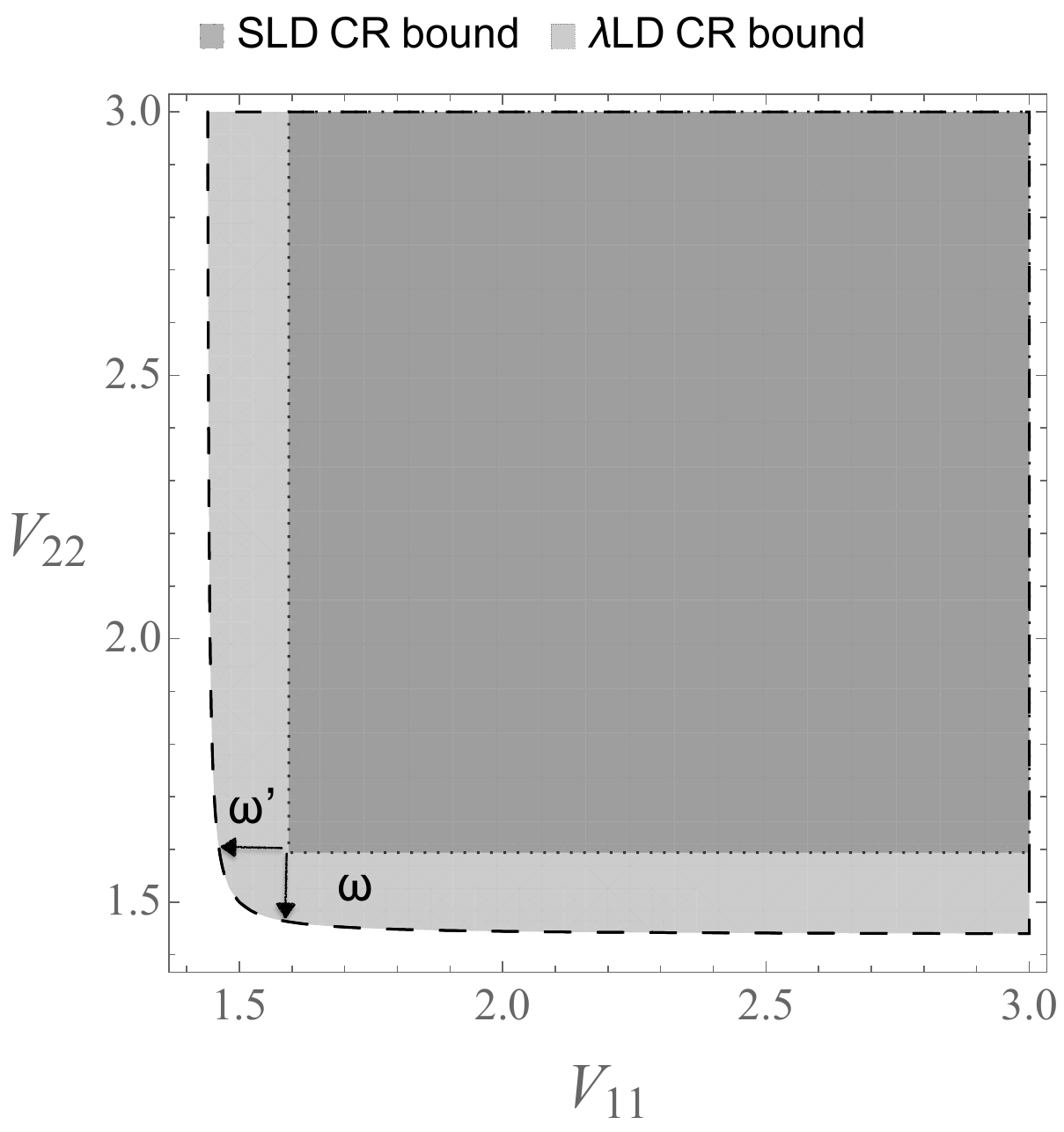}
\caption{The case of $\omega < 0$ 
and $\omega'<0$.} The intersection of the SLD and the $\lambda$ bounds 
(darker gray) is the same as the SLD bounds. 
\label{fig_app2}
\end{center}
\end{figure}

Suppose we have obtain the $\lambda$LD CR inequality for the MSE matrix: $\sfV\ge J^{-1}_{\: \lambda}$. 
Note that $\lambda=0$ case corresponds to the SLD CR inequality as a special case. 
Let $\sfV_{ii}$ ($i=1,2$) be the diagonal components of $\sfV$, 
then the SLD CR inequality gives two independent inequalities 
\be
\sfV_{ii}\ge J^{-1}_{\: \mathrm{S} , ii}\quad(i=1,2). 
\ee
The allowed region for $(\sfV_{11},\sfV_{22})$ is represented by an upper half square region 
on the $\sfV_{11}\sfV_{22}$ plane. 
For example, we plot this region by the light gray region in Fig.~\ref{fig_app1}. 
Next, we fix a particular value of $\lambda\neq0$ and consider the matrix inequality 
$\sfV-J^{-1}_{\: \lambda}\ge0$. 
By taking the determinant of this inequality, we have 
\begin{multline}
(\sfV_{\:11} - J^{-1}_{\: \lambda , 11})(\sfV_{\:22}  - J^{-1}_{\: \lambda , 22})
-\big(\sfV_{12}-\mathrm{Re}(J^{-1}_{\: \lambda , 12})\big) \big(\sfV_{21}-\mathrm{Re}(J^{-1}_{\: \lambda , 21} 
)\big)\\
- | \mathrm{Im} (J^{-1}_{\: \lambda , 12}) |^2\ge0. 
\end{multline}
This inequality implies that the diagonal components, which are of our interest, 
must satisfy the following relation. 
\be
(\sfV_{\:11} - J^{-1}_{\: \lambda , 11})(\sfV_{\:22}  - J^{-1}_{\: \lambda , 22}) 
\ge | \mathrm{Im} (J^{-1}_{\: \lambda , 12}) |^2. 
\label{eq:lambdaLD_ineq}
\ee
A similar derivation of Eq.~\eqref{eq:lambdaLD_ineq} is given in Appendix A of \cite{sf2}. 
We now have two possibilities when considering two regions allowed by the SLD and the $\lambda$LD CR inequalities. 
The first case is when the boundary for the inequality Eq.~\eqref{eq:lambdaLD_ineq} 
intersects the boundary lines of the SLD CR inequality. The first case is shown in Fig.~\ref{fig_app1}. 
The other case is when there is no intersection between two boundaries. 
This second case is shown in Fig.~\ref{fig_app2}. 

In the first case (the SLD and the $\lambda$LD CR bounds have intersections), 
we can narrow the region where the diagonal components of the MSE matrix $\sfV_{11}$ and $\sfV_{22}$. 
The allowed region by the two CR inequalities is shown by the dark grey region in Fig.~\ref{fig_app1}. 
In this case, we conclude the existence of a tradeoff relation. 
For the second case, on the other hand, one cannot obtain any useful information from the two CR inequalities, 
since the SLD CR bound completely dominates the $\lambda$LD CR bound. 
We shall define the indicator $\omega$ and $\omega'$
by the length of the lines BA and BA' in Fig.~\ref{fig_app1}, respectively. 
A graphical explanation of the indicators $\omega$ 
and $\omega'$ are given in Fig.~\ref{fig_app1} and Fig.~\ref{fig_app2}. 
When the indicator $\omega$ is positive, we can confirm that a tradeoff relation exists. 
To put our idea into the equation, note that the boundary given by the $\lambda$LD CR bound is expressed as
\be
(\sfV_{\:11} - J^{-1}_{\: \lambda , 11})(\sfV_{\:22}  - J^{-1}_{\: \lambda , 22}) 
= | \mathrm{Im} (J^{-1}_{\: \lambda , 12}) |^2. 
\label{eq:lambdaLD_boundary}
\ee
Whereas the boundary given by the SLD which is the dotted line in Fig.~\ref{fig_app1} is expressed as
\be
\sfV_{\:11} = J^{-1}_{\: \mathrm{S} , 11}, \quad \sfV_{\:11} = J^{-1}_{\: \mathrm{S} , 22}. 
\ee
The $\sfV_{22}$ component of the intersection A (Fig.6) is given by the $\sfV_{\:22}$ at 
$J^{-1}_{\: \mathrm{S} , 11}$.
Hence, we have
\begin{align}
\left. \sfV_{\:22} \right|_{\sfV_{\:11}=J^{-1}_{\: \mathrm{S} , 11}}
&=\frac{ | \mathrm{Im} J^{-1}_{\: \lambda , 12} |^2}{J^{-1}_{\: \mathrm{S} , 11} - J^{-1}_{\: \lambda , 11}}+J^{-1}_{\: \lambda , 22}. 
\label{eq:sl_boundary1} 
\end{align}
As can be seen in Fig.~\ref{fig_app1}, 
an explicit expression of the point A is 
A=$(J^{-1}_{\: \mathrm{S} , 11}, J^{-1}_{\: \mathrm{S} , 22})$.
We can define the indicator $\omega$ by subtracting 
$J^{-1}_{\: \mathrm{S} , 22}$ from the right hand side of \eqref{eq:sl_boundary1}. 
\begin{align} 
\omega = \frac{| \mathrm{Im} J^{-1}_{\: \lambda , 12}  |^2 - 
 (J^{-1}_{\: \mathrm{S} , 11}-J^{-1}_{\: \lambda , 11} )(J^{-1}_{\: \mathrm{S} , 22}-J^{-1}_{\: \lambda , 22})}
 {J^{-1}_{\: \mathrm{S} , 11}-J^{-1}_{\: \lambda , 11}}. \label{eq:omega}
\end{align} 
The $\sfV_{11}$ component of the intersection A' (Fig.6) is given by the $\sfV_{\:11}$ at 
$J^{-1}_{\: \mathrm{S} , 22}$.
We have
\begin{align}
\left. \sfV_{\:11} \right|_{\sfV_{\:22}=J^{-1}_{\: \mathrm{S} , 22}}
&=\frac{ | \mathrm{Im} J^{-1}_{\: \lambda , 12} |^2}{J^{-1}_{\: \mathrm{S} , 22} - J^{-1}_{\: \lambda , 22}}+J^{-1}_{\: \lambda , 11}. 
\label{eq:sl_boundary2}
\end{align}
By a similar consideration, $\omega'$ is obtained as
\begin{align} 
\omega' = \frac{| \mathrm{Im} J^{-1}_{\: \lambda , 12}  |^2 - 
 (J^{-1}_{\: \mathrm{S} , 11}-J^{-1}_{\: \lambda , 11} )(J^{-1}_{\: \mathrm{S} , 11}-J^{-1}_{\: \lambda , 11})}
 {J^{-1}_{\: \mathrm{S} , 22}-J^{-1}_{\: \lambda , 22}}. 
\end{align} 
Since the numerators of the indicators $\omega$ and $\omega'$ are the same, 
if $\omega$=0 holds, $\omega'$ holds, and vice versa.
\section{Proof of $\Theta=(1- \zeta^2)/\xi  > 1$} \label{sec:det_Jlambda}
This section provides the proofs of the inequalities used in this paper. 
Note that the integral $\xi$ is always positive, then $\Theta=(1- \zeta^2)/\xi  > 1$ implies the following inequality. 
\be
1- \zeta^2 > \xi\ \Rightarrow\ 1- \zeta^2 >0. 
\ee
We first show the following relation between $\xi$ and $\zeta$. 
\be
\xi + \frac{\sqrt{2}\zeta}{\kappa^\prime\v} 
= 1 + \frac{\sqrt{\pi}}{2 \kappa^\prime} \e^{{\kappa^\prime}^2} \erfc(\kappa^\prime), \label{eq:xi_eta1}
\ee
where $\kappa^\prime = m \kappa$, and $\erfc(x)$ denotes the complementary error function.

This is shown by first substituting explicit expressions of $\xi$, Eq.~\eqref{eq:zeta} and $\zeta$, Eq.~\eqref{eq:xi} 
into the left hand side of Eq.~\eqref{eq:xi_eta1}. 
Then, the definition of the complementary error function gives  
\begin{align}
\xi + \frac{\sqrt{2} \zeta}{\kappa^\prime \v} 
 &= {\kappa^\prime}^2 \int^\infty_0 dt  2t \sqrt{1+t^2} \e^{-{\kappa^\prime}^2 t^2} \\ 
 &= 1 + \frac{\sqrt{\pi}}{2 \kappa^\prime} \e^{{\kappa^\prime}^2} \erfc(\kappa^\prime). \label{eq:zeta_xi} \quad \square
\end{align}
Next, we show that $\Theta=(1- \zeta^2)/\xi >1$ holds for any $\kappa > 0$ and $0<\v \leq 1$.
\subsubsection{Proof}
Let us define $T$ by
\be
T= 1- \zeta^2 - \xi.
\ee
Then, $T > 0$ is a necessary and sufficient condition for $(1- \zeta^2)/{\xi} >1$, 
since $\xi$ is always positive by definition. 
To show $T>0$, we show $\zeta^2+\xi < 1$ instead because 
$T>0  \iff \zeta^2+\xi < 1$. 

By the definition of $\zeta$, Eq.~\eqref{eq:zeta}, 
\begin{align}
\zeta&={\sqrt{2}} \v \int^\infty_0 dt \frac{ {\kp}^3 t^3 \e^{-{\kp}^2 t^2} }{\sqrt{1+t^2} 
 + \sqrt{1- {\v}^2}},  
 \end{align}
we obtain the following inequality for $\zeta$. 
\be
{\sqrt{2}} \v \int^\infty_0 dt \frac{ {\kp}^3 t^3 \e^{-{\kp}^2 t^2} }{\sqrt{1+t^2} + 1}
\le  \zeta  \le
{\sqrt{2}} \v \int^\infty_0 dt \frac{ {\kp}^3 t^3 \e^{-{\kp}^2 t^2} }{\sqrt{1+t^2}}.
 \nonumber 
\ee
The integrations in the inequality above are explicitly written as
\begin{align}
 \int^\infty_0 dt \frac{ {\kp}^3 t^3 \e^{-{\kp}^2 t^2} }{\sqrt{1+t^2} + 1}
&=\frac{\sqrt{\pi}}{4} \e^{{\kappa^\prime}^2} \erfc(\kappa^\prime), 
\label{eq:upper}  \\
\int^\infty_0 dt \frac{ {\kp}^3 t^3 \e^{-{\kp}^2 t^2} }{\sqrt{1+t^2}}
&=  \frac{\kappa^\prime}{2} + \frac{\sqrt{\pi}}{4} \e^{{\kappa^\prime}^2} (1 - 2 {\kappa^\prime}^2) \, \erfc(\kappa^\prime) . 
\label{eq:lower}
\end{align}
Then, we have the following inequalities regarding $\zeta^2$ and $\xi$
\begin{align}
\zeta^2 &\leq \frac{\pi}{8} V \e^{2 {\kp}^2} (\erfc(\kp))^2 \leq \frac{\pi}{8} \e^{2 {\kp}^2} (\erfc(\kp))^2, \\
\xi
& \leq -\frac{\sqrt{2}}{\kp } [\frac{\kappa^\prime}{\sqrt{2}} + \frac{\sqrt{2\pi}}{4} \e^{{\kappa^\prime}^2} (1 - 2 {\kappa^\prime}^2) \, \erfc(\kappa^\prime) ] \nonumber\\
&\qquad+1+ \frac{\sqrt{\pi}}{2 {\kp}} \e^{{\kp}^2}\erfc(\kp) \nonumber \\
&=\sqrt{\pi} \kp \e^{{\kp}^2} \, \erfc(\kp). \label{eq:xi_inequality}
\end{align}
We use Eq.~\eqref{eq:xi_eta1} to derive Eq.~\eqref{eq:xi_inequality}. 
Therefore, we obtain an inequality for $\zeta^2+\xi$, 
\begin{align}
\zeta^2+\xi 
&\leq \frac{\pi}{8} \e^{2{\kp}^2} (\erfc(\kp))^2
+\sqrt{\pi} \kp \e^{{\kp}^2} \, \erfc(\kp).
\end{align}
We can show that the right-hand side of the inequality above is a monotonically increasing function of $\kp$. 
Next, we check if $\zeta^2+\xi \leq 1$ holds when $\kp \gg 1$. 
By using the asymptotic expansion of the complementary error function $\erfc(x)$,
\be
\erfc(x)=\frac{\e^{-x^2}}{\sqrt{\pi} x} \sum_{n=0}^{\infty} (-1)^n \frac{(2n-1)!!}{2^n x^{2n}}. \nonumber 
\ee
For $\kappa^\prime \gg 1$, we have
\begin{align}
\frac{\pi}{8} \e^{2{\kp}^2} (\erfc(\kp))^2 &= \frac{1}{8 {\kp}^2}+\mathcal{O}(\frac{1}{{\kp}^4}), \nonumber \\
\sqrt{\pi} \kp \e^{{\kp}^2} \, \erfc(\kp) &= 1 - \frac{1}{2 {\kp}^2}+\mathcal{O}(\frac{1}{{\kp}^4}).
\end{align}
Thus, for $\kappa^\prime \gg 1$, we obtain
\be
\zeta^2+\xi < \frac{\pi}{8} \e^{2{\kp}^2} (\erfc(\kp))^2
+\sqrt{\pi} \kp \e^{{\kp}^2} \, \erfc(\kp) = 1- \frac{3}{8 {\kp}^2} + \mathcal{O}(\frac{1}{{\kp}^4}).
\ee
The inequality $\zeta^2+\xi < 1$ holds for any $\kp >0$ and $V \in$ (0,1], where $\kp= m \kappa$. 

Therefore,
\be
 \Theta=\frac{1- \zeta^2}{\xi} > 1, 
\ee
is proven. $\square$

\section{Indicator $\omega$ and the solution of $\omega(\lambda, \kappa, {\v})=0$} 
In this section, we give supplemental materials regarding our method to confirm the existence of the tradeoff relation. 
We give the properties of the indicator $\omega$. 
\subsection{$\omega$: a decreasing function of $\lambda$ for any given $\kappa$ and $\v$} \label{sec:fuction_omega}
We show that  
$\omega(\lambda, \kappa, {\v})$ is a monotonically decreasing function of $\lambda$ for any $\kappa >0$ and
any $\v \in (0,1]$.
\subsubsection{Proof}
For any given $\kappa >0$ and for any given $\v \in (0,1]$,
the first derivative of the right hand side of Eq.~\eqref{eq:explicit_omega2} with respect to $\lambda$ is 
calculated as 
\be
\frac{\partial \omega(\lambda, \kappa, {\v})}{\partial \lambda}
= - \frac{\lambda(a \lambda^4 + b \lambda^2+c)}
{ \{ \zeta ^4 - \zeta ^2 [ (1- \lambda ^2 ) \xi ^2+2 ]-\lambda ^2 \xi ^2+1 \}^2}, \label{ch7_eq:deriv_omega}
\ee
where
\begin{align*}
a&= \xi ^4 \left(1-\zeta ^2 \right)^2,  \\
b&=- 2 \xi ^2 (1- \zeta ^2 ) \left[ ( 1-  \zeta ^2 )^2-\xi^2 \zeta ^2\right], \\
c&=  -2\sqrt{2} \zeta ^6 \left(2 \xi ^2+1\right)+\zeta ^4 \left(\xi ^4+5 \xi ^2+3\right)
-4 \zeta ^2 \left(\xi ^2+1\right)+1.
\end{align*}
From these, we are to show the first derivative of $\omega(\lambda, \kappa, {\v})$
with respect to $\lambda$ is negative. 
We just need to check if $a \lambda^4 + b \lambda^2 + c > 0$ holds  
since the denominator is positive. 

We show here that the inequality $a s^2 + b s +c >0$, where $s = \lambda^2$. 
By definition, $s \in (0,1)$. Since $a >0$, 
 $a s^2 + b s +c >0$ holds if and only if its discriminant is negative, i.e., $D < 0$. 
The discriminant $D$ is explicitly written as 
\be
D= - 4 \zeta ^2 \xi ^6 (1- \zeta ^2 )^3 (\Theta^2-1) .
\ee
As shown in Appendix~\ref{sec:det_Jlambda}, $\Theta=(1-\zeta ^2)/\xi > 1$ and $\Theta>1$ hold, 
and hence $D < 0$ holds, i.e., $a \lambda^4 + b \lambda^2 + c > 0$ ($a > 0$) holds. 
Hence, the first derivative of $\omega(\lambda, \kappa, {\v})$ with respect to $\lambda$, 
the right hand side of Eq.~\eqref{ch7_eq:deriv_omega} is always negative. 
The indicator $\omega(\lambda, \kappa, {\v})$ is a strictly monotonically decreasing function of $\lambda$ 
for any given $\kappa >0$ and for any given $\v \in (0,1]$. $\square$
\subsection{Positivity of $\omega(0, \kappa, \v)$}\label{sec:limit0}
Let us define a limit $\omega(0, \kappa, {\v})$ by 
\be
\omega(0, \kappa, {\v}) = \lim_{\lambda \to 0} \omega(\lambda, \kappa, {\v}).
\ee
From Eq.~\eqref{eq:explicit_omega2}, we have  
\begin{align}
\omega(0, \kappa, {\v}) 
&= \frac{ \kappa ^2 \zeta ^4 }{ 2\xi \Theta ( \Theta ^2-\zeta ^2 )},  \label{ch7_eq:omega0}
\end{align}
where 
\be
\Theta = \frac{1- \zeta ^2}{\xi}.
\ee
As given in Appendix~\ref{sec:det_Jlambda}, $\Theta > 1$ and 
$1- \zeta ^2 > 0$ hold. We have $\Theta -\zeta ^2 > 0$ using these inequalities.  
Therefore, 
\be
\omega(0, \kappa, {\v}) > 0, 
\ee
holds for any $\kappa > 0$ and $\v \in (0,1]$. 
We remark that the case of $\v=0$ (the rest frame) is excluded. 

\subsection{Negativity of $\omega(1, \kappa, \v)$}\label{sec:limit1}
Next, we investigate the other limit, $\lambda \rightarrow 1$. 
Let us define a limit $\omega(1, \kappa, {\v})$ by 
\be
\omega(1, \kappa, {\v}) = \lim_{\lambda \to 1} \omega(\lambda, \kappa, {\v}).
\ee
From the explicit expression of $\omega(\lambda, \kappa, {\v})$, i.e., Eq.~\eqref{eq:explicit_omega2}
$\omega(1, \kappa, {\v})$ is expressed as
\begin{align}
\omega(1, \kappa, {\v})
&=- \frac{\kappa^2}{2 (1-  \zeta^2)}.
\end{align}
As shown in Appendix~\ref{sec:det_Jlambda}, $1- \zeta^2 > 0$ holds. 
Hence, we have 
\be
\omega(1, \kappa, {\v}) < 0,
\ee
for any $\kappa > 0$ and $\v \in (0,1]$. 

\subsection{Solution $\lambda^\star$ for $\omega(\lambda, \kappa, \v)=0$} \label{sec:lambda_star_sol}
In the following, we are to derive the solution for $\omega(\lambda, \kappa, {\v})=0$. 
To be precise, let us also check the denominator of Eq.~\eqref{eq:explicit_omega2}. 
Beside the trivial factor ${\kappa^2}/{2 (1- \zeta^2)}$, it is expressed as 
\be
 \xi^2\left[ \lambda ^2 (1- \zeta ^2)+ \zeta ^2  -\frac{(1-\zeta ^2)^2}{\xi^2}\right].
\ee
The first term $\lambda ^2 (1-\zeta ^2)+ \zeta ^2$ is evaluated as
\be
\lambda ^2 (1-\zeta ^2)+\zeta ^2 < 1- \zeta ^2 +  \zeta ^2 = 1.
\ee
We use $\lambda^2 < 1$. We also know $\Theta=(1-\zeta ^2 )/ \xi > 1$.
Therefore, the denominator of the right-hand side of Eq.~\eqref{eq:explicit_omega2} is always negative. 

Therefore, $\omega(\lambda, \kappa, {\v})=0$ holds, if and only if 
the numerator is zero. We then need to solve 
\begin{align}
\lambda ^2 (1- \zeta ^2 )^2 - \xi ^2 [ \lambda ^2 (1-  \zeta ^2)+\zeta ^2 ]^2=0.
\label{secG:omega}
\end{align}
By factoring the numerator Eq.~\eqref{secG:omega}, we obtain an equivalent condition 
for $\omega(\lambda^\ast, \kappa, {\v}) = 0$ as follows. 
\begin{align}
& \{ (1-\zeta ^2 ) \lambda^\ast + \xi [ (\lambda^\ast)^2 (1 - \zeta^2 ) + \zeta^2 ] \} \nonumber \\
&\times
\{ (1-\zeta^2 ) \lambda^\ast - \xi  [( \lambda^\ast) ^2 (1 - \zeta^2 ) + \zeta^2 ]  \} = 0. \label{eq:inequ1}
\end{align}
From $ 1- \zeta ^2  > 0$ and $ 0 < \xi \leq 1$ for any $\kappa >0$ and $\v \in (0,1]$, the following inequality holds.
\be
(1-\zeta ^2 ) \lambda^\ast + \xi [ (\lambda^\ast )^2 (1 - \zeta^2 ) +  \zeta^2] > 0.
\ee
Therefore, Eq.~\eqref{eq:inequ1} reduces to
\begin{align*}
&(1-\zeta ^2 ) \lambda^\ast - \xi [ (\lambda^\ast) ^2 (1 - \zeta^2 ) +  \zeta^2 ] = 0\\ 
\Leftrightarrow\ & \xi(1 - \zeta^2 )(\lambda^\ast) ^2-(1-\zeta ^2 ) \lambda^\ast+\xi\zeta^2=0.
\end{align*}

The solutions of the equation, $\lambda^\ast_\pm$ is given by 
\begin{align}
\lambda^\ast_\pm
&=\frac{1}{2\xi} \left(1 \pm \sqrt{1 - \frac{4\xi^2\zeta ^2}{ 1- \zeta ^2 }} \, \right) .
\end{align} 
In the limit of $\kappa \rightarrow 0$ when $\v=1$, $\xi$ approaches zero. 
Hence, $\lambda^\ast_+ |_{\v=1}$ diverges. 
As we know that there is a unique solution, we take the $\lambda^\ast_-$ as the solution. 
\begin{align}
\lambda^\ast
&=\frac{1}{2\xi} \left(1 - \sqrt{1 - \frac{4 \xi^2\zeta ^2}{ 1-\zeta ^2 }} \, \right) . \quad \square
\end{align} 

\section{Relativistic limit}\label{sec:rel_limit}
In this Appendix, we analyze the relativistic limit ($\v=1$). 
First, we have
\begin{align}
\zeta
&= {\sqrt{2}}{\v} \int^\infty_0 dt \frac{ {\kappa^\prime}^3 {t}^3 \e^{-{\kappa^\prime}^2 {t}^2}}{\sqrt{1+{t}^2} 
 + \sqrt{1- {\v}^2}},  
\end{align}
where $\kappa^\prime = m \kappa$. 
By setting $\v=1$ and the use of the definition of the complementary error function, 
we obtain
\begin{align}
\zeta_{\rm rel} &={\sqrt{2}}
  \int^\infty_0 dt \frac{ {\kappa^\prime}^3 {t}^3 \e^{-{\kappa^\prime}^2 {t}^2}}{\sqrt{1+{t}^2}}\\
&= \frac{\kappa^\prime}{\sqrt{2}} + \frac{\sqrt{2\pi}}{4} \e^{{\kappa^\prime}^2} (1 - 2 {\kappa^\prime}^2) \, \erfc(\kappa^\prime) . 
\label{eq:zeta_rel}
\end{align}
The other function $\xi$ in the relativistic limit is also obtained as 
\be\label{eq:xi_rel}
\xi_{\rm rel}=\sqrt{\pi}\kappa^\prime\e^{{\kappa^\prime}^2}\erfc(\kappa^\prime). 
\ee

We can easily verify that $\zeta_{\rm rel}$ is a monotonically decreasing function of $\kappa^\prime$ 
by the property of the complementary error function. 
We can also show the following limits: 
\begin{align}
\lim_{\kappa^\prime\to0}\zeta_{\rm rel}&=\frac{\sqrt{2\pi}}{4},\\
\lim_{\kappa^\prime\to\infty}\zeta_{\rm rel}&=0.
\end{align}
Similarly, $\xi_{\rm rel}$ is a monotonically increasing function of $\kappa^\prime$ 
and its limits are 
\begin{align}
\lim_{\kappa^\prime\to0}\xi_{\rm rel}&=0,\\
\lim_{\kappa^\prime\to\infty}\xi_{\rm rel}&=1.
\end{align}


\end{document}